\documentclass[preprint2]{aastex}
\pdfoutput=1

\usepackage{amsmath, amssymb}
\usepackage[]{movie15}
\usepackage{epsfig}
\usepackage{morefloats}
\usepackage{graphicx}
\usepackage{color}
\usepackage{hyperref}
\hypersetup{
linktocpage=true,
pdfborderstyle={/S/S/W 1},
hyperindex=true,
bookmarks=true,
bookmarksopen=true,
bookmarksnumbered=true}

\newcommand \mdotbin {\dot{M}(q \ne 0)}
\newcommand \mdotsig {\dot{M}(q = 0)}
\newcommand \beq {\begin{equation}}
\newcommand \enq {\end{equation}}
\newcommand \omgb {\Omega_{\rm bin}}

\newcommand \rin {r_{\rm in}}
\newcommand \rout {r_{\rm out}}
\makeatletter

\newcommand{\Rmnum}[1]{\expandafter\@slowromancap\romannumeral #1@}
\makeatother

\slugcomment{}

\shorttitle{3D MHD Simulations of Circumbinary Disks}
\shortauthors{Shi, \& Krolik}

\begin{document}


\title{Three Dimensional MHD Simulation of Circumbinary Accretion Disks --\textrm{2}. 
Net Accretion Rate}


\author{Ji-Ming Shi\altaffilmark{1} and Julian H. Krolik\altaffilmark{2}}
\altaffiltext{1}{Department of Astrophysical Sciences, Princeton University, 4 Ivy Lane,
    Princeton, NJ 08544}
\altaffiltext{2}{Department of Physics and Astronomy, Johns Hopkins
University, Baltimore, MD 21218}

\begin{abstract}
When an accretion disk surrounds a binary rotating in the same sense, the binary exerts
strong torques on the gas.   Analytic work in the 1D approximation indicated that these
torques sharply diminish or even eliminate accretion from the disk onto the binary.   However,
recent 2D and 3D simulational work has shown at most modest diminution.    We present
new MHD simulations demonstrating that for binaries with mass ratios of 1 and 0.1 there is
essentially no difference between the accretion rate at large radius in the disk and the accretion
rate onto the binary.   To resolve the discrepancy with earlier analytic estimates, we identify
the small subset of gas trajectories traveling from the inner edge of the disk to the binary and
show how the full accretion rate is concentrated onto them as a result of stream-disk shocks
driven by the binary torques.
\end{abstract}


\keywords{accretion, accretion disks --- binaries: general --- MHD --- methods:
numerical}

\section{INTRODUCTION \label{sec:intr}}

\begin{deluxetable}{cccccc}
\tablecolumns{6} \tablewidth{0pc}
\tablecaption{Properties of Accretion Disk Simulations}
\tablehead{\colhead{Label} & \colhead{Type of Simulation} &\colhead{$q$}  & \colhead{Resolution\tablenotemark{a}
} & \colhead{Radial Extent\tablenotemark{b}} &
\colhead{$\langle\dot{M}(r_{\rm in})\rangle_t\tablenotemark{c}~(\sqrt{GMa}\Sigma_0)$} 
}
\startdata
S2DE        & Hydrodynamics & 0.0   & $256\times64$           & $(0.8,40) $  & ...\\
S3DEQ       & MHD           & 0.0   & $512\times400\times96$  & $(0.8,40) $  & $0.0081$\\
S3DE        & MHD           & 0.0   & $512\times400\times384$ & $(0.8,40) $  & $0.0085$\\
B3DE        & MHD           & 1.0   & $512\times400\times384$ & $(0.8,40)$   & $0.011-0.014$\\
B3DEq       & MHD           & 0.1   & $480\times400\times384$ & $(1.02,40)$  & $0.013-0.017$\\
\enddata 
\small{\tablenotetext{a}{Cell counts are listed as $r\times\theta\times\phi$ for 3D MHD simulations and
$r\times\theta$ for 2D hydrodynamic simulations.}}
\small{\tablenotetext{b}{In units of $a$. The azimuthal extent is $(0,2\pi)$ for all MHD simulations
except S3DEQ, which is $(0,\pi/2)$.}}
\small{\tablenotetext{c}{Time1averaged accretion rate measured at the inner boundary.  The accretion
rate in the quarter-disk simulation S3DEQ is multiplied by 4.  For binary runs, time averages at
both early and late stages are presented and separated by a dash.}}
\label{tab:runs}
\end{deluxetable}

Binary systems often form within a gaseous disk environment.   Because the
tidal torques exerted by the binary on the disk are repulsive when the disk and the
binary orbit in the same sense, it has long been thought that these torques severely limit,
or perhaps entirely prevent, accretion from the disk onto the binary \cite{Pringle91}.
As a result, a very low-density cavity forms within $\sim 2a$ of the binary
center of mass, where $a$ is the binary's semi-major axis.
Despite this prediction, observations indicate accretion
onto such binary systems at a level comparable to their single star 
counterparts.  This is clearly detected for low-mass binary stars in nearby star-forming 
regions \citep[e.g.,][]{WG2001}.  High velocity gas flows are observed bridging 
gaps that are believed to be cleared out by protoplanetary companions 
\citep[e.g.,][]{Casassus2013,Rosenfeld2014}. There are even detections of accretion 
onto planetary mass companions \citep[e.g.,][]{Bowler2011,Zhou2014}. 
A growing number of dual AGN candidates have also been reported
\citep[e.g.,][]{Komossa2003,Rodriguez2006,Comerford2011}; if systems like these
become mutually bound, they could become accreting binaries.
It is therefore important to understand how gas is able to accrete despite these torques,
both to be able to use electromagnetic signatures \citep{Roedig2014} as diagnostics and
also to learn how accretion influences the evolution of these systems.

Much work has already been expended investigating how the binary torque reshapes the
surrounding circumbinary disk \citep[e.g.,][]{al94,cmm2006,mm08,shi12,N12}.   However,
how the torque regulates the gas accretion is a less well-developed subject.     The 1D
analysis of \cite{Pringle91} has been extended, but adopting that paper's assumption
of zero accretion \citep[e.g.,][]{mp05,lodato09,Liu10}.    More recently, there have been
a number of 2D and 3D simulations of such systems
\citep[e.g.,][]{al96,bryden99,mm08,dotti09,shi12,N12,dorazio13,farris2014}.
These consistently find the cavity predicted analytically, but also
accretion at a rate comparable to or somewhat smaller
than in the outer disk.   Typically, the accretion takes place in narrow, high velocity streams
emanating from the edge of the cavity and spiraling inward toward the central binary.
Moreover, these studies have shown significant consequences of the accretion for the
development of the system.
In protoplanetary disks, the amount of accretion through the gap controls
the mass growth of a gap-clearing planet \citep[][]{bryden99,lubow99,dlb06}
and the surface density within the gap \citep[][]{zhu2011,fung2014}. 
The angular momentum associated with the advecting gas also strongly influence the orbital evolution of
the binary \citep[][]{shi12,roedig12}. 

It is therefore crucial to quantify the net accretion fraction ($\epsilon$) of a prograde circumbinary disk, i.e. the
ratio between the time-averaged net accretion rate $\mdotbin$ and the rate that would take place
in the absence of binary torque $\mdotsig$:
\beq
\epsilon \equiv \frac{\mdotbin}{\mdotsig} .
\enq
For the extreme small mass-ratio case (we define the mass-ratio $q \equiv M_2/M_1 \leq 1$),
previous studies using viscous hydrodynamics simulations have found that there is a threshold
mass-ratio $q\sim 10^{-3}$, such that for smaller $q$ the accretion rate is nearly unchanged
when compared to that for a single point-mass case; i.e. $\epsilon \simeq 1$ when $q \lesssim 10^{-3}$.
For values of $q$ closer to unity, the results so far are somewhat mixed.   Most simulations have found that
$\epsilon$ is reduced by a factor of a few \citep[][]{bryden99,lubow99,shi12,N12,dorazio13},
but the dependence on $q$ remains poorly defined.   For example, \citet{dorazio13} saw
diminished accretion for $q\gtrsim 0.01$ binaries, while \citet[][]{farris2014} saw
no signs of suppression for $0.026 < q < 1$.   The situation is further clouded by the fact that
previous work (except for \cite{shi12} and \cite{N12}) has used the approximation of 2D hydrodynamics,
in which internal accretion stresses, both within the accretion disk itself and in the streams
traversing the cavity, are described by the phenomenological $\alpha$ viscosity model, rather
than the actual physical mechanism, correlated MHD turbulence.

In this paper, we carry out 3D MHD simulations to investigate accretion in circumbinary disks
around equal mass ($q=1$) and unequal mass ($q=0.1$) binaries.   We assume the disk and binary
are coplanar, orbit in the same sense, and the binary orbit is a circle with fixed semi-major axis $a$.
For a control experiment, we also simulate their single point-mass counterpart.    Comparison
between the single mass run and the circumbinary runs provides the answers to two key
questions: 
(1) To what extent does the accretion rate of a binary differ from that of a single object, i.e.,
what is the value of $\epsilon$? 
(2) How does the accretion rate from circumbinary disks depend on the mass ratio $q$ of the binary,
i.e., how does $\epsilon$ vary with $q$? 
As we will show, the answer to the first question is also the answer to the second: $\epsilon \simeq 1$
for all $q$.    This answer leads
to a further question that we will also attempt to answer: what was missing from the early 1D analytic
studies that led them to conclude $\epsilon \simeq 0$?

We organize this paper as follows: In \S~\ref{sec:numerical}, we describe the
physical model and numerical setup of our single point-mass and circumbinary disk simulations.
In \S~\ref{sec:result}, we present our simulation results.  We analyze these results physically in
\S~\ref{sec:analysis}.   Finally, we summarize our conclusions and
discuss the implications of our findings in \S~\ref{sec:conclusion}.

\begin{figure*}[]
\centering{
\includegraphics[width=8cm]{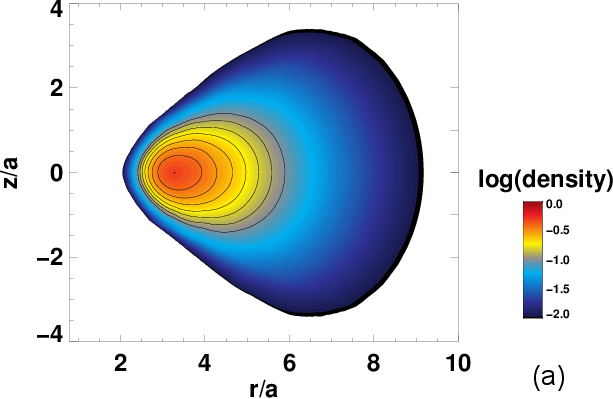} \\
\includegraphics[width=18cm]{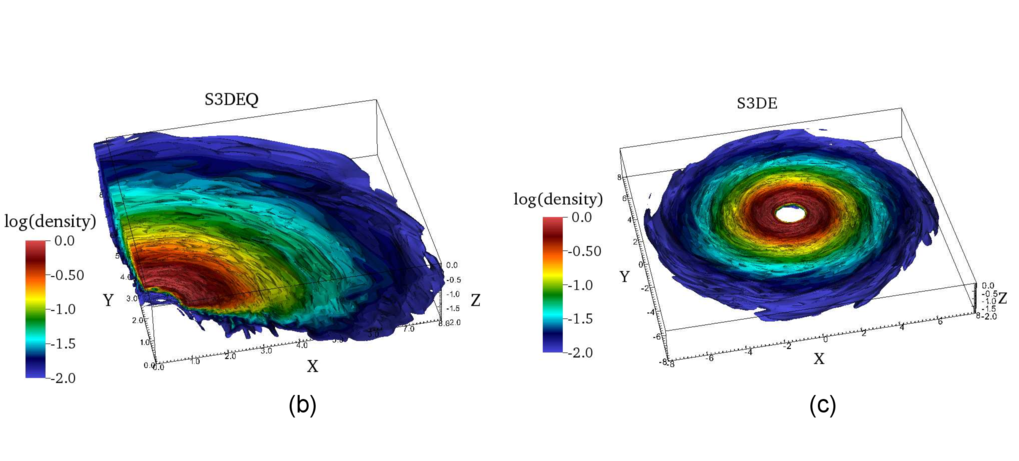}
}
\caption{\small{(a)~Density distribution (color) at the end of the 2D hydrodynamical
simulation S2DE, which is used to provide initial conditions for the 3D MHD quarter disk run S3DEQ.
Note that the vertical scale is stretched relative to the horizontal.
The black contour lines mark the initial magnetic field loops. (b) ~Density isosurfaces (color) of the
bottom half disk of S3DEQ at $t=800\omgb^{-1}$.  This data is used as the initial conditions of S3DE after
combining four identical copies to make the whole $2\pi$ disk. (c) ~ Density isosurfaces
(color) of the bottom half disk of S3DE at $t=1000\omgb^{-1}$.   This data becomes the initial
conditions for our circumbinary simulations. 
}}
\label{fig:initial}
\end{figure*}

\section{Numerical Simulations\label{sec:numerical}}
In this section, we discuss details about the numerical experiments we carried out in order to explore
the effects of the binary torques on the accretion rate.  We used the same code as
\citet{shi12}, which is a modern version of the 3D, time-explicit Eulerian finite-differencing ZEUS
code for MHD \citep{stone92a,stone92b,hs95}.

\subsection{Model Setup\label{sec:setup}}
The model setup is very similar to the one described in \citet{shi12}.   Our account here is therefore
very brief and emphasizes the few differences between our new simulations and the one presented
in our earlier paper.  We also summarize the main properties of all runs in this paper in
Table~\ref{tab:runs} for reference.

We construct our disks in an inertial frame with origin at the center of mass.  We set the gravitational
constant $G$ and the central mass $M$ to be unity, whether it is a single object or a binary. The binary
separation $a=1$ is the simulation lengthscale\footnote{In the $q=0$ simulation, $a$ has no physical
interpretation other than the code-unit of length.}. The time unit is therefore $\omgb^{-1}=\sqrt{a^3/(GM)}$.
The density is normalized to the initial midplane value $\rho_0=1$ and the surface density unit is defined
to be $\Sigma_0=\rho_0 a$.  
Similar to \citet{shi12}, we adopt a globally isothermal equation of state with a
fixed sound speed, but we set $c_s=0.1$, twice that of the previous paper.   A larger sound speed allows
us to achieve better resolution and also perform longer duration simulations. Again, we assume the
disk mass is considerably smaller than the mass of the central object and therefore neglect the self-gravity of
the disk.  

As discussed in \citet[][see section~2.2]{shi12}, the simulation grid needs to resolve three
different length scales: the disk scale height $H$, the maximum growth-rate wavelength of the MRI $\lambda_{\rm
MRI}=8\pi/\sqrt{15}v_A/\Omega(R)$, and the spiral density wavelength $\lambda_{\rm d}\sim 2\pi c_s
/\omgb$, where $v_A$ is the Alfven speed and $\Omega(R)$ is the disk rotational frequency. 
We constructed our grid in spherical coordinates following the same scheme as in \citet{shi12},
which was originally proposed by \citet{N10}:  logarithmic in the radial direction, uniformly
spaced in azimuthal angle, and spaced according to a polynomial function in polar angle in order
to concentrate cells near the orbital plane (Equation~(1) of \citet{shi12}, with $\xi=0.825$, $\theta_c=0.1$
and $n=9$). There were $[N_{\rm r},400,384]$ cells in $(r,\theta,\phi)$, covering a computational
domain spanning $[\rin, \rout]$ radially, $[\theta_c, \pi-\theta_c]$ meridionally and $[0,2\pi]$
azimuthally, where $\rout=40a$, $\theta_c=0.1$, $\rin$ is the radius of the circular central
cut-out ($0.8a$ for $q=0$ and $q=1$, increased to $1.02a$ for $q=0.1$ to
confine the secondary inside the excision), and $N_{\rm r}=512$
for $q=0$ and $q=1$, diminished to $480$ for $q=0.1$ run as a result of the larger $\rin$.
In terms of minimum cell size, our polar angle resolution is a factor $\sim 1.4$ better than in
\citet{shi12}. With a doubled disk scaleheight ($c_s=0.1$ vs. $c_s=0.05$), we therefore
have about twice as many cells per scaleheight as in \citet{shi12}. In recent years, standards have
been developed for achieving resolution adequate to describe the principal features of MRI-driven MHD
turbulence: at least 10--20 cells per $2\pi v_{Az,\phi}/\Omega$, where $v_{Az,\phi}$ is the Alfven
speed associated with the vertical ($z$) component of the magnetic field or azimuthal ($\phi$) component
\citep{HGK11,HRGK13}.   We typically have $\simeq 20$~cells per characteristic vertical wavelength and at
least 10~cells per characteristic azimuthal wavelength.

We choose strict outflow boundary conditions for the inner and outer radial surfaces and also at the
edges of the polar axis cutout; all inward velocities are set to zero on these boundaries. Periodic boundary
conditions are used for the azimuthal
direction. For the magnetic field in the radial and meridional directions, we set the transverse components of
the field to be zero in the ghost zones.  The components normal to the boundaries are
calculated by imposing the divergence-free constraint.


\subsection{Numerical Experiments\label{sec:experiments}}

\begin{figure}[h!]
\epsscale{0.5}
\plotone{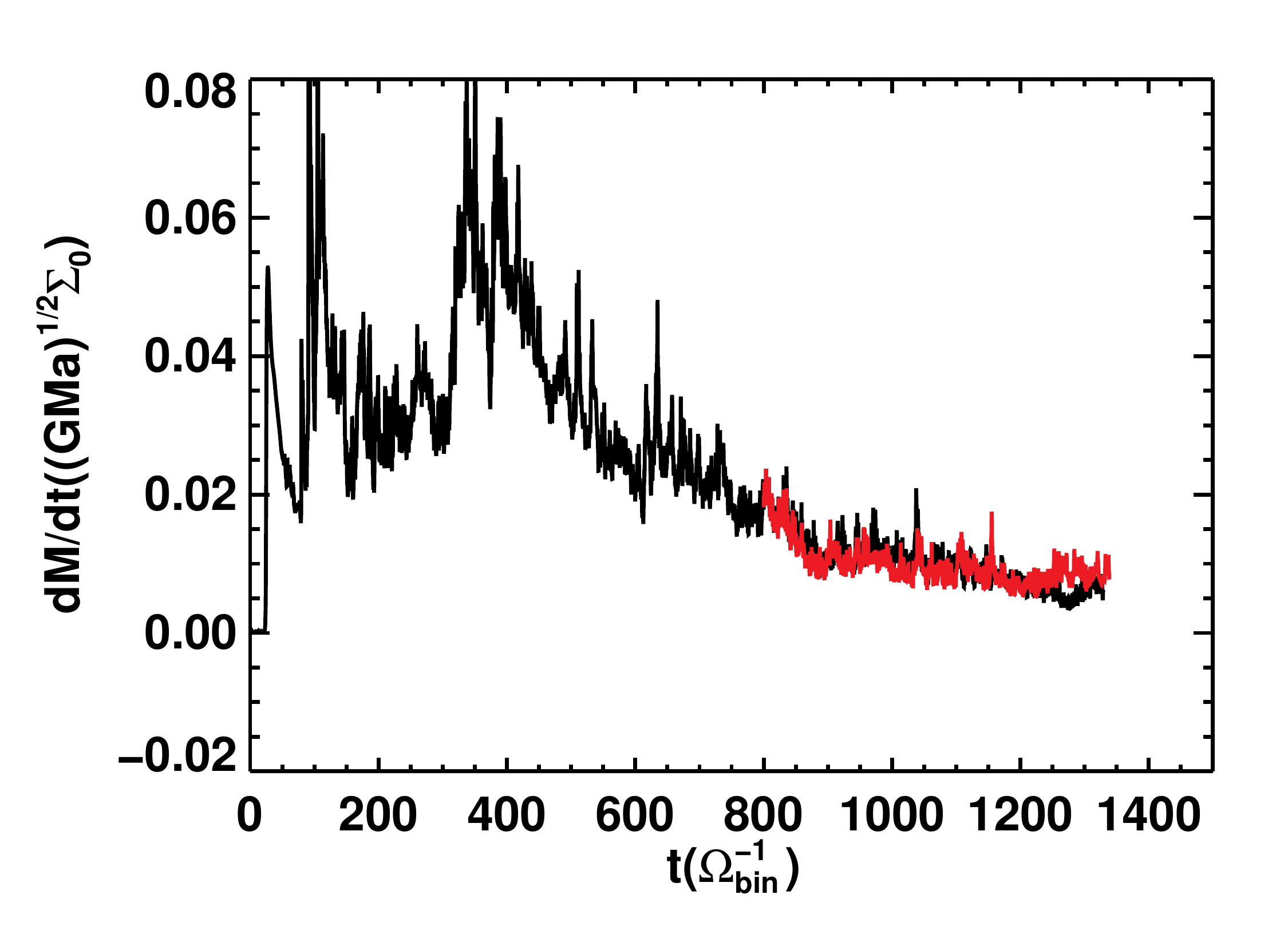}
\plotone{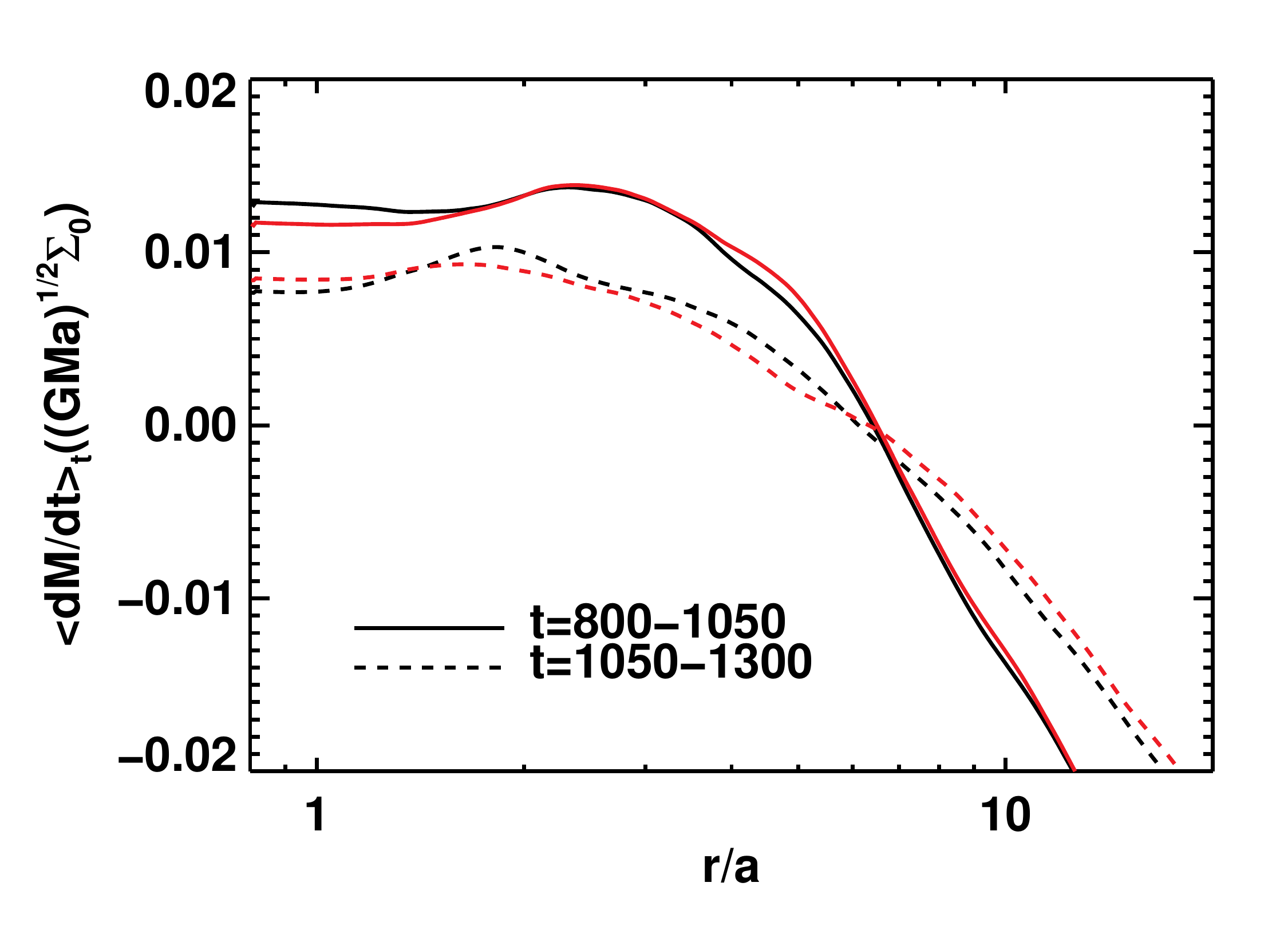}
\plotone{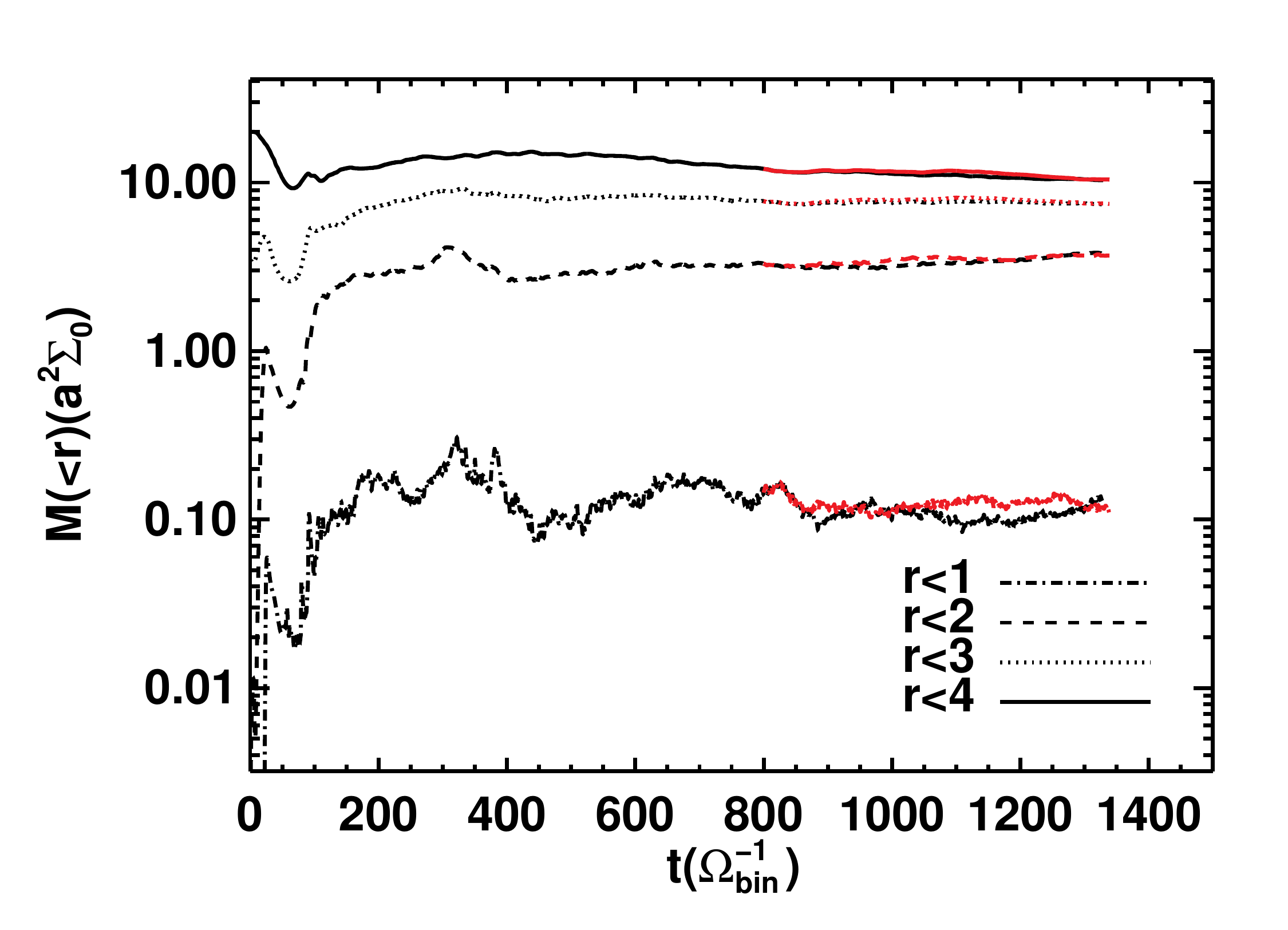}
\caption{\small{Accretion history of the single mass quarter-circle simulation S3DEQ (black) and the
                full-circle S3DE MHD (red) disks. \textsl{Top}:~Inner edge
		accretion rate as a function of time. \textsl{Middle}:~Time-averaged accretion
		rate as a function of radius. \textsl{Bottom}:~Mass interior to a given
		radius as a function of time.
		The accretion rates and enclosed mass for the quarter-disk are multiplied by
		$4$ for comparison with whole disk values. 
		Note the very similar behaviors between the quarter disk and the
		complete disk, in agreement with previous simulations: the time averaged accretion
                of MHD turbulent disks is nearly independent of the extent of the azimuthal domain if
                it is more than $\pi/3$ \citep{hawley2000,PN2003}.
}}
\label{fig:s3de}
\end{figure}

\subsubsection{Single Mass Run: $q=0$\label{sec:num_q=0}}
The single mass simulation both serves as a reference point and provides the initial conditions for the circumbinary
simulations discussed later. 
In order to speed up these lengthy simulations, we built a three-step ladder:

\textbf{Step~1. 2D Hydrodynamic Disk:} 
We start from a 2D axisymmetric, inviscid, hydrodynamic simulation (S2DE) in the $(r,\theta)$ plane with the same
numerical and physical parameters as described above except that the parameter $\xi$ in the $\theta$-grid
definition is 0.92 and the number of cells is only lower $256\times 64$ in $r \times \theta$.
The goal of this 2D simulation is to get rid of initial transients and to find a
quasi-equilibrium disk solution.
The initial configuration follows that of \citet[][section~2.3]{shi12}:  the disk stretches from
$r=3a$ to $r=6a$ with constant midplane density $\rho_0$ and is vertically
hydrostatic.   We evolve the disk for $2000\omgb^{-1}$, when the disk reaches a steady state
(see Figure~\ref{fig:initial}(a)). 

\textbf{Step~2. 3D MHD Quarter Disk:}
We then use the final dump from S2DE as the initial conditions for a $\pi/2$ wedge of a 3D MHD
disk (S3DEQ). The initial disk for S3DEQ is therefore axisymmetric.   Its initial density and angular velocity
are interpolated from the results of S2DE data.   Motions in other directions are neglected as they are
very small. 
The initial contours of magnetic vector potential $A_\phi$ are a set of nested poloidal loops following
the contours of the density within the main body of the disk (the contour lines in
Figure~\ref{fig:initial}(a)).    The values of the $A_{\phi}$ contours are $A_0(\rho - 0.1\rho_0)$, where $A_0$ is
a constant determined by requiring the average plasma $\beta=100$. The magnetic field $\mathbf{B}$ is computed
by taking the curl of the vector potential.   
Run S3DEQ begins at $t=0$ and lasts until $t=1330 \omgb^{-1}$.  A snapshot of the quarter-disk at
$t=800\omgb^{-1}$ is shown in Figure~\ref{fig:initial}(b). We find the quarter
disk approaches
steady accretion for $r < 5a$ between $t=800\omgb^{-1}$ and $1300\omgb^{-1}$ (see
Figure~\ref{fig:s3de}).

\textbf{Step~3. 3D MHD full Disk:}
We then patch together four identical copies of the $\pi/2$ disk from S3DEQ at $t=800\omgb^{-1}$ to build the
initial conditions for a full $2\pi$ disk (S3DE). This run continues from $t=800\omgb^{-1}$ through
$1340\omgb^{-1}$. The evolution history (see Figure~\ref{fig:s3de}), nearly the same as the quarter
disk run, indicates that by $\sim 800\omgb^{-1}$ the full disk undergoes
quasi-steady accretion\footnote{The time averaged accretion rate at
different radii suggests our $q=0$ disk is approaching inflow equilibrium for $r\lesssim 4$-$5a$
(see the second panel in Figure~\ref{fig:s3de}).   Outside that radius, the accretion rate gradually
shrinks, changing sign to outflow at $r \sim 7a$.    A small amount of mass outflow at large radius
carries the angular momentum transported from inside.}.
We show a clipped isosurfaces plot of this fully
turbulent disk at $t=1000\omgb^{-1}$ in Figure~\ref{fig:initial}(c).   The disk at this moment serves
as the initial conditions for the binary runs that are discussed next.

\subsubsection{Binary Runs: $q \neq 0$\label{sec:num_q.ne.0}}

We performed two binary simulations, one with mass ratio $q=1$(B3DE) and one with $q=0.1$(B3DEq). 
For both runs, we use S3DE data at $t=1000\omgb^{-1}$ as the initial conditions.   Both also retain
the domain size, numerical resolution, and physical parameters of the $q=0$ run; the only change
is to replace the point-mass at the center with either an equal-mass binary or a $q=0.1$ binary with the
same total mass.   In both binary cases, the orbit is constant and circular, and
rotates counter-clockwise, prograde with respect to the disk.

Everything about the $q=0.1$ case is the same except for the extent of the radial grid.   Because the secondary
is $\simeq 0.9a$ from the center of mass, we increase the
inner excision size from $0.8a$ to $\rin=1.02a$. This change results in a truncation of the initial
disk of the single mass run, but we keep all other aspects, such as the resolution and outer boundary of
the domain, intact.

\section{Results\label{sec:result}}

\begin{figure} [t!]
\epsscale{0.7}
\plotone{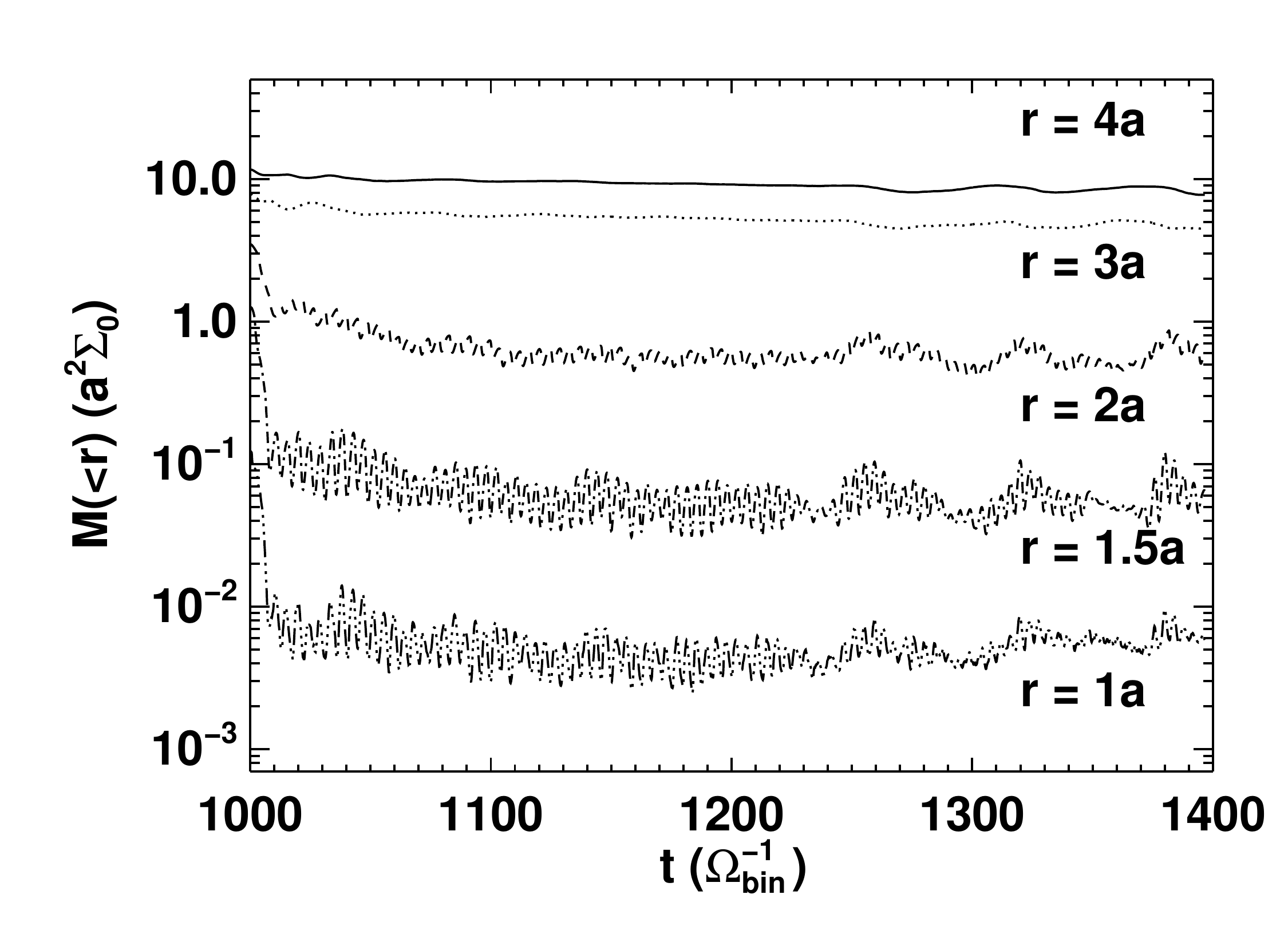}
\plotone{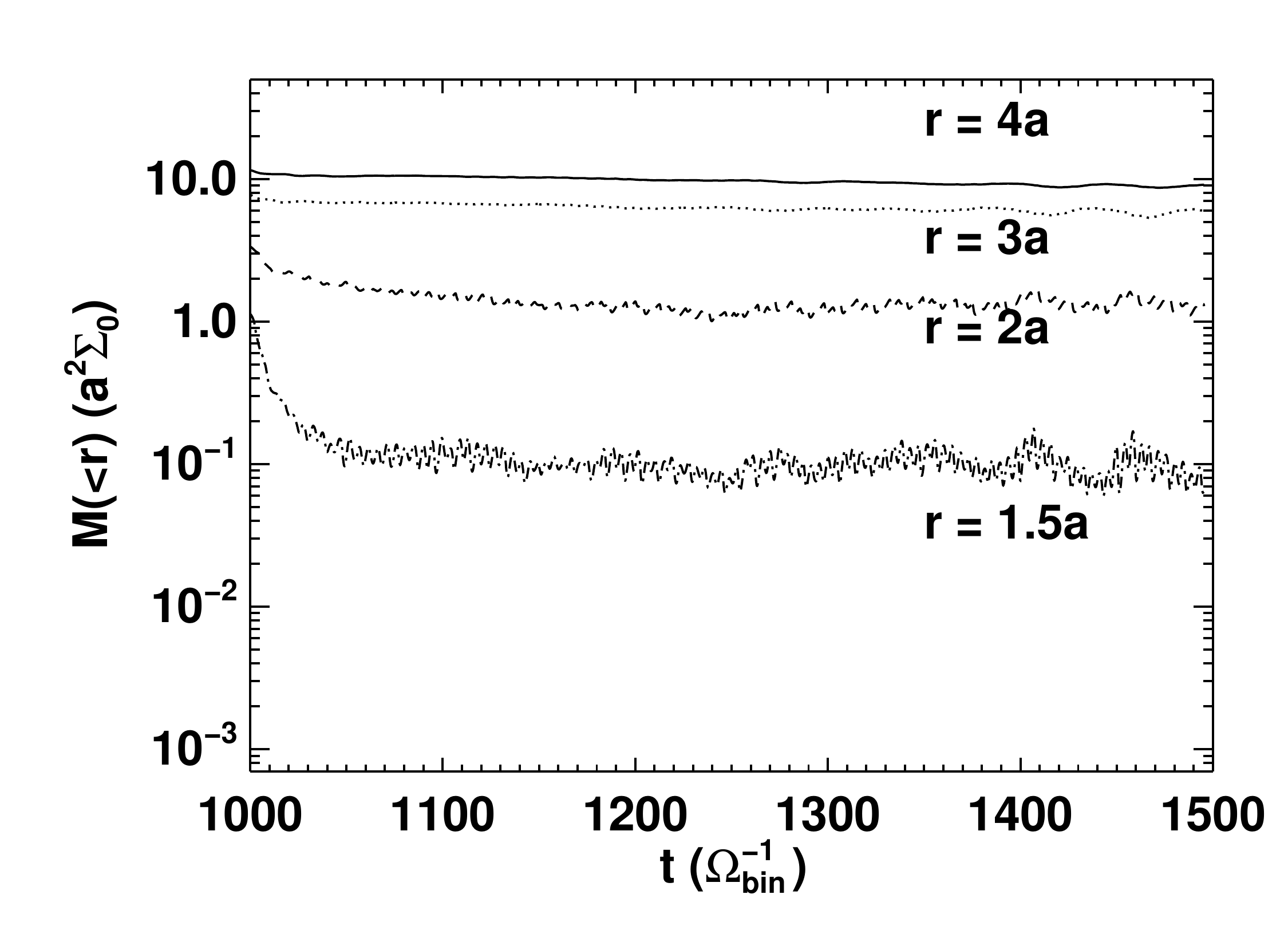}
\caption{\small{ ~History of disk mass interior to a given radius for $q=1$ (top) and $q=0.1$
		circumbinary disks.
}}
\label{fig:mass_r_t}
\end{figure}
\subsection{$q=1$ Run \label{sec:res_q=1}}
\begin{figure}[!ht]
\epsscale{0.7}
\plotone{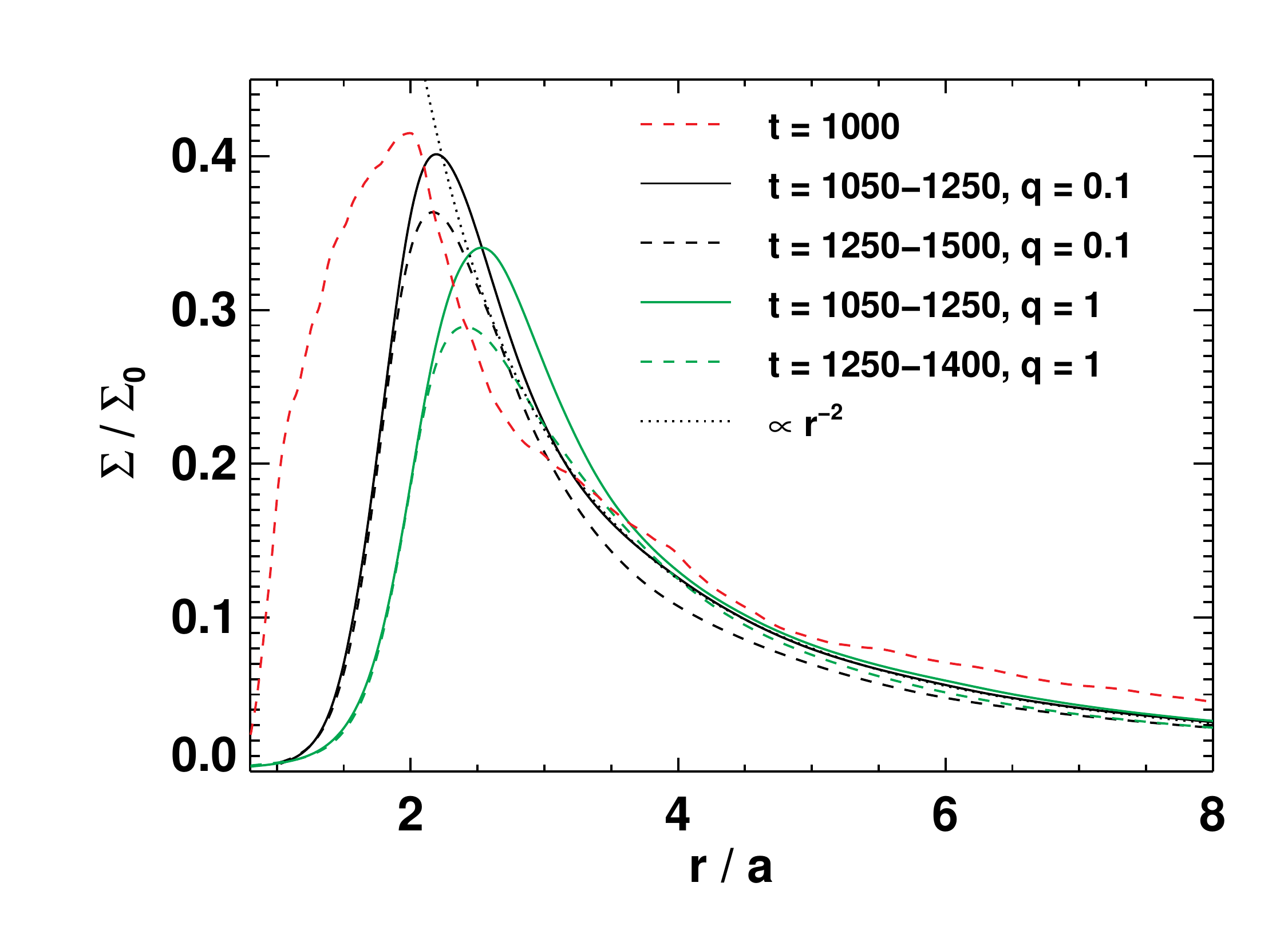}
\caption{\small{Surface densities for the initial condition (red dashed) and the time-averaged
                $q=0.1$ (black) and $q=1$ (green) cases.   Solid curves are early-time, dashed 
		curves are late-time averages. Solids are for the early time averages and dashed 
		for the late time averages. For reference, a $\propto r^{-2}$ scaling is shown 
		by the black dotted curve. }}
\label{fig:sigma_q}
\end{figure}

The initial disk adjusts to the new potential within $1$--$2$ binary periods after the equal-mass binary replaces
the point-mass.
During this initial transient phase, the binary clears out a cavity around itself in which the density is
$10^{-3}$--$10^{-2}$ of what it was when there was a point-mass at the center.   A small amount of mass initially
found at $r \lesssim 2a$ passes through the inner boundary (about $2\%$ of the total disk mass), 
but most of the disk mass within $r < 2a$ is pushed outward by the binary torque.  After $\gtrsim 50\omgb^{-1}$, the
circumbinary disk gradually reaches a quasi-steady state.  In its inner portion ($r \lesssim 4a$), the mass
interior to $r$ becomes nearly constant in time (first panel in Figure~\ref{fig:mass_r_t}) so that
the surface density's radial profile also changes only very slowly (Fig.~\ref{fig:sigma_q}).
We also note that the late-time average surface density of the outer disk follows the $\Sigma \propto r^{-2}$
scaling observed in \citet{shi12} quite well.

\begin{figure}[h!]
\epsscale{0.7}
\plotone{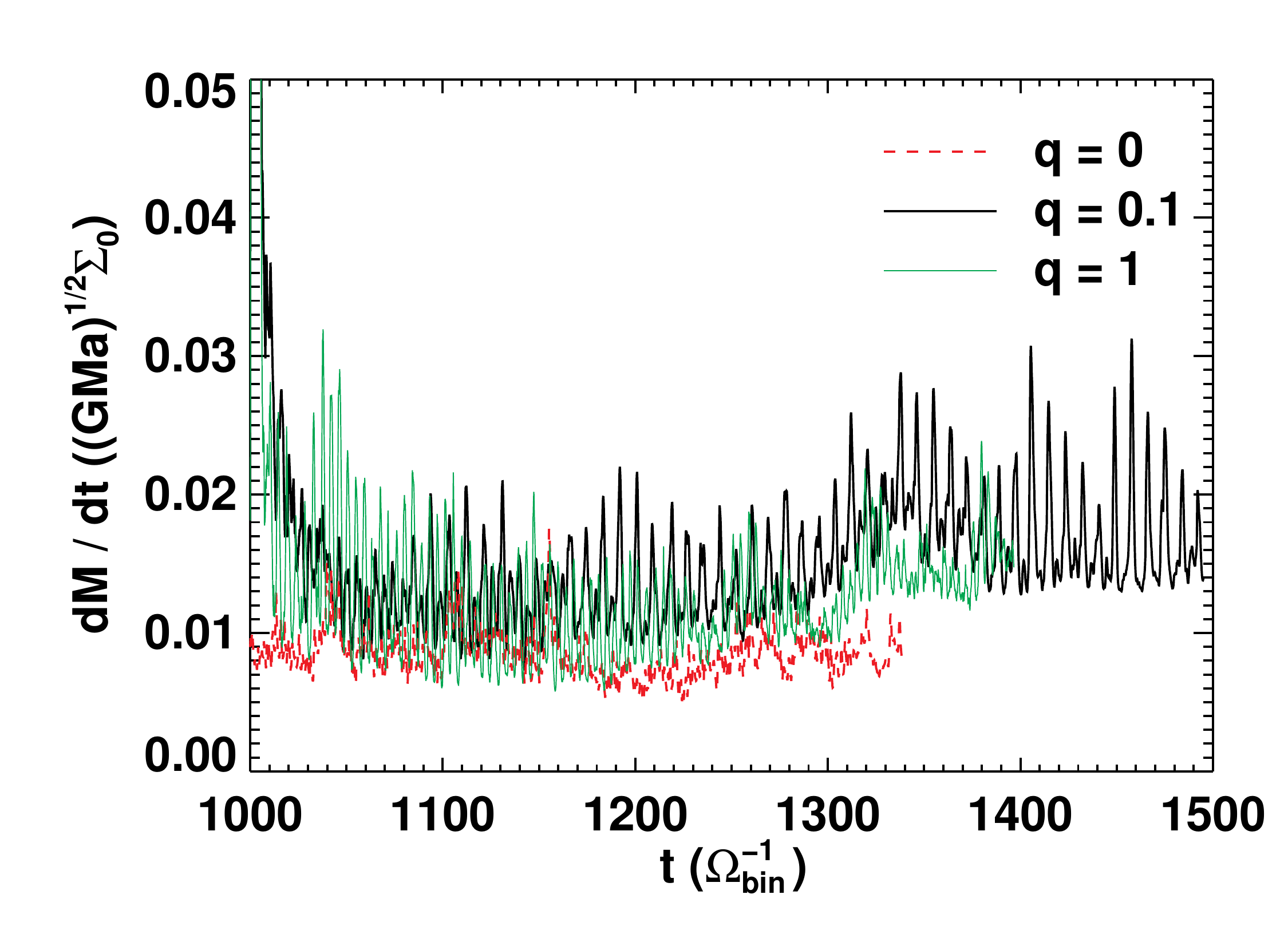}
\plotone{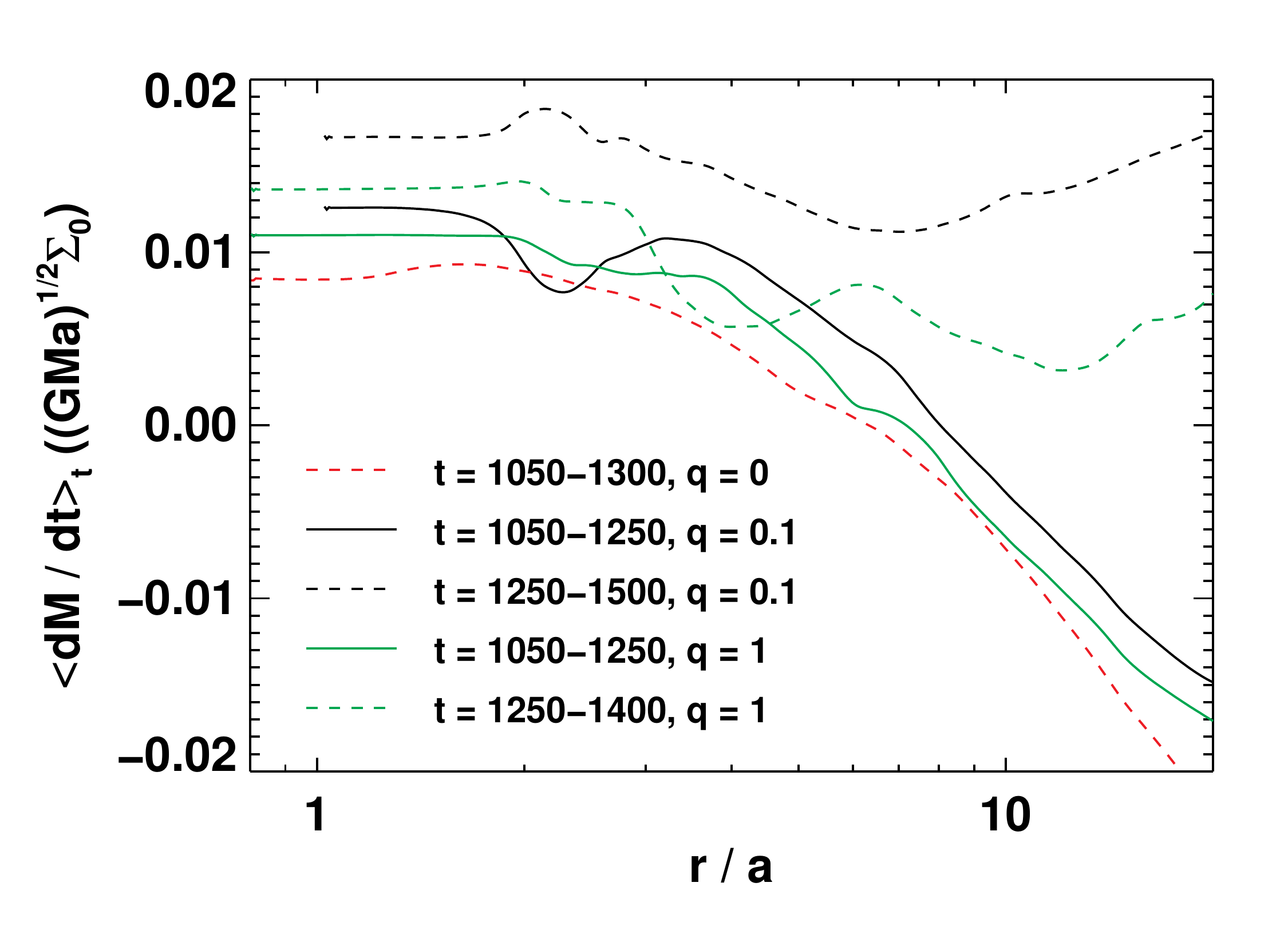}
\caption{\small{\textsl{Top}:~Accretion rates through the inner boundary of $q=0$ (red dashed), 
		$q=0.1$ (black), and $q=1$ (green) disks. \textsl{Bottom}:~Time-averaged accretion 
		rates for $q=0$ (red dashed), $q=0.1$ (black), and $q=1$ (green) as a function of 
		radius.  For the $q\neq 0 $ cases, both early (solid) and late (dashed) time averages 
                are presented. }}
\label{fig:mdot_t_r}
\end{figure}

After the first $\sim 50\omgb^{-1}$, the accretion histories show that the $q=1$ disk
has also reached a quasi-steady state with respect to this property (Fig.~\ref{fig:mdot_t_r}).
Clearly, the accretion is not
hampered by the central binary (contrast the red dashed curve for the $q=0$ disk with the green curve for
$q=1$).   Although
the strong binary torque clears out a low density cavity near the center, the cavity is not empty of gas.
Streams of gas still flow through the potential maxima at the $L2$ and $L3$ points (see Fig.~\ref{fig:avgmdot})
and maintain accretion at a rate comparable to the single mass case.
The overall time-averaged accretion rate of the $q=1$ disk is in fact a bit greater than that of the $q=0$ case:
$\dot{M} (r=r_{\rm in})\simeq 0.011 (GMa)^{1/2}\Sigma_0$ over $t=1050$--$1250\omgb^{-1}$ (`early-time') and
$\simeq 0.014 (GMa)^{1/2}\Sigma_0$ over $1250$--$1400\omgb^{-1}$ (`late-time'),
compared to $0.0085 (GMa)^{1/2}\Sigma_0$ of the $q=0$ disk (from $t=1000\omgb^{-1}$
until the end of the simulation).   These numbers translate to an accretion efficiency $\epsilon \simeq 1.3$--$1.6$. 
This result is in approximate agreement with the viscous hydrodynamics simulations of \citet{dorazio13},
in which they found the accretion rate of an equal mass binary is $\sim 0.93$ times the single-mass case
(see their Table~3).    It is in even better agreement with the results of \citet{farris2014}, who found a ratio
of 1.55.

We show the radial dependence of the time-averaged accretion rate $\dot{M}(r)$ in
Figure~\ref{fig:mdot_t_r}. For both early ($t=1050$-$1250\omgb^{-1}$) and late time averages
($t=1250$--$1400\omgb^{-1}$), $\dot{M}(r)$ for both $q=1$ and $q=0.1$ exceeds that of the
$q=0$ disk at all radii. The $q=1$ disk achieves a fairly steady accretion within
$r\lesssim 4a$, quite similar to a single-mass disk. At late times, the accretion rate at 
smaller radii increases by $\sim 20\%$, and the zone of nearly constant accretion radius
as a function of radius extends outward from $r \simeq 4a$, characteristic of early times,
to $r \simeq 25a$.

\begin{figure}[t!]
\epsscale{0.7}
\plotone{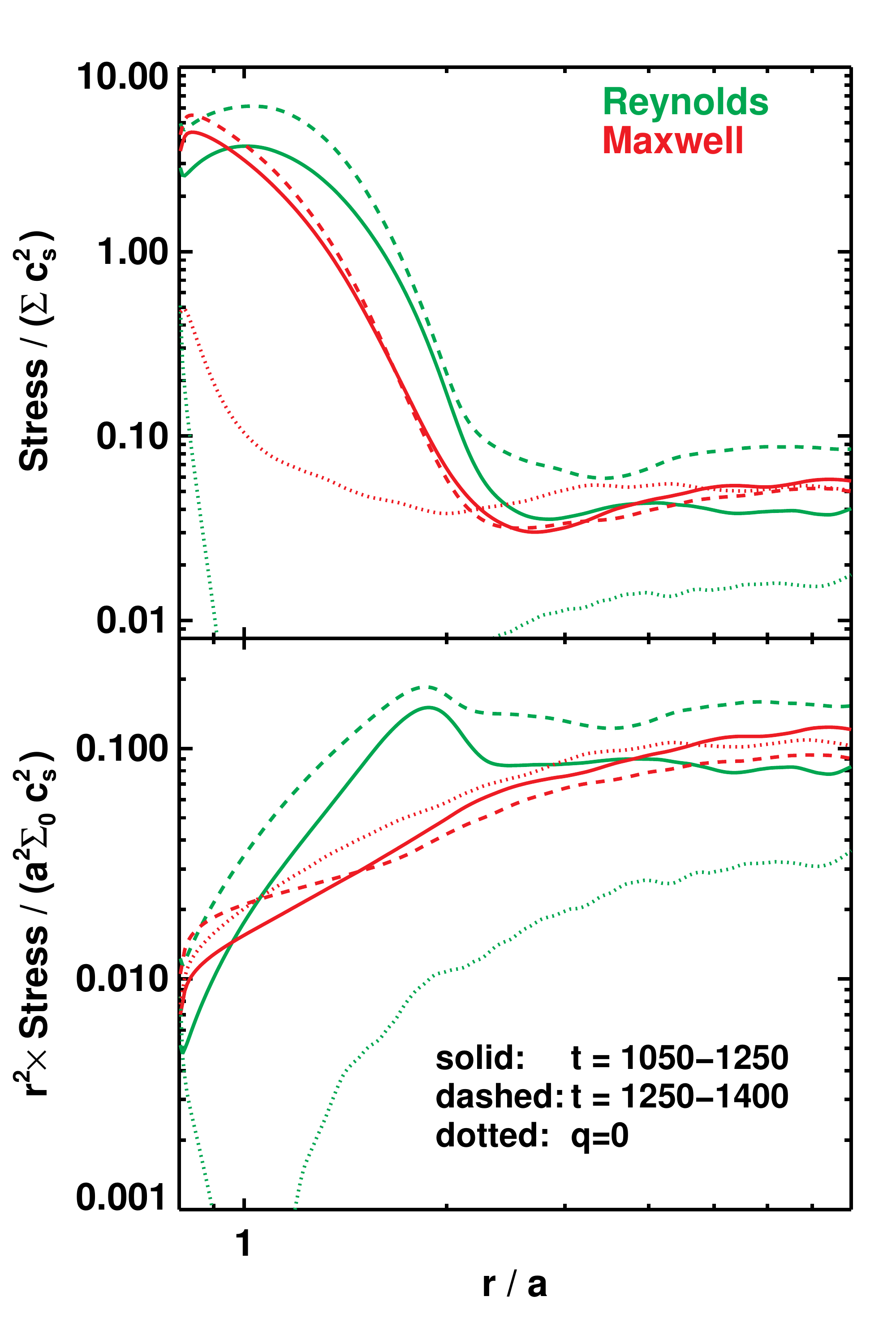}
\caption{\small{ Time averaged stress to pressure ratios (top) and their associated angular 
		momentum fluxes for the $q=1$ disk.   Different line styles denote different 
		time spans: $t=1050$-$1250$ (solid), $t=1250$-$1400$ (dashed), and 
		$t=1050$-$1300$ of the $q=0$ run (dotted);  different colors distinguish 
		Reynolds (green) and Maxwell (red) stresses. There is an increase of Reynolds 
		stress and its corresponding angular momentum flux at later times, owing to 
                the $m=1$ spiral waves.}}
\label{fig:str_q=1}
\end{figure}

The different levels of accretion observed in Figure~\ref{fig:mdot_t_r} are directly connected to
variations in the internal stresses.   As shown in Figure~\ref{fig:str_q=1}, although
the Maxwell stress changes little with time, the Reynolds stress rises by a factor of $\sim 2$
after $t=1250\omgb^{-1}$ at all radii outside $r\simeq 2a$.  This increase in internal stress then
drives a larger accretion rate at late time.

\begin{figure*}[ht!]
{\hbox{\hspace{-0.3cm}
\includegraphics[width=15cm]{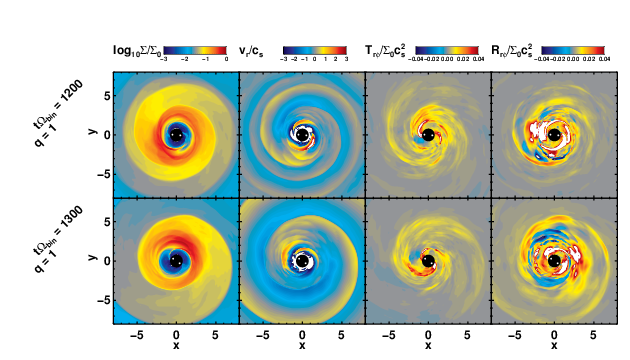}}}
\caption{\small{Four disk diagnostics before and after the spiral wave phase transition at 
		$\sim 1250\omgb^{-1}$ in the $q=1$ simulation. 
		\textsl{Upper row}: $t=1200\omgb^{-1}$, \textsl{lower row}: $t=1300\omgb^{-1}$.   
		From left to right, the variables shown are: surface density, density weighted 
		radial velocity (normalized to the sound speed), vertically integrated Maxwell 
		and Reynolds stresses in units of $\Sigma_0 c_s^2$. The two white dots in the 
		central cut-out region represent the binary members. White indicates off-scale 
                high values. }}
\label{fig:twophase_q=1}
\end{figure*}

The increase in Reynolds stress occurs at the same time ($\sim 1250\omgb^{-1}$) as the disk's
spiral density waves change from a tightly-wrapped $m=2$ pattern to a single-armed wave
(Fig.~\ref{fig:twophase_q=1}).  Although the spiral waves have little effect on the Maxwell stress, the
vertically-integrated Reynolds stress is enhanced along the density crest of the $m=1$ wave.  For instance,
the region of large Reynolds stress between $r=4a$ and $8a$ at $t=1300\omgb^{-1}$ (the yellow
spiral arm winding from 9 o'clock to 6 o'clock) is absent in the $t=1200\omgb^{-1}$ snapshot.   It
appears that the single-armed spiral wave is more effective at conveying angular momentum
outward than the two-armed wave.

\subsection{$q=0.1$ Run \label{sec:res_q=0.1}}
The accretion flow in the $q=0.1$ disk is quite similar to that seen in the $q=1$ case. The disk reaches its
quasi-steady state after the first $\sim 50 \omgb^{-1}$. The history of enclosed disk mass within a given
radius in Figure~\ref{fig:mass_r_t} indicates a slowly evolving steady state in the inner part of
the circumbinary disk. 
%
Further evidence that an approximate state of inflow equilibrium has been reached for $r < 4a$
is given by the time-averaged radial profiles of $\dot{M}(r)$ (Figure~\ref{fig:mdot_t_r}).   Also like
the $q=1$ case, the accretion rate for $q=0.1$ gradually increases over the course of the simulation,
rising by $\simeq 30\%$.

\begin{figure*}[ht!]
{\hbox{\hspace{-0.3cm}
\includegraphics[width=15cm]{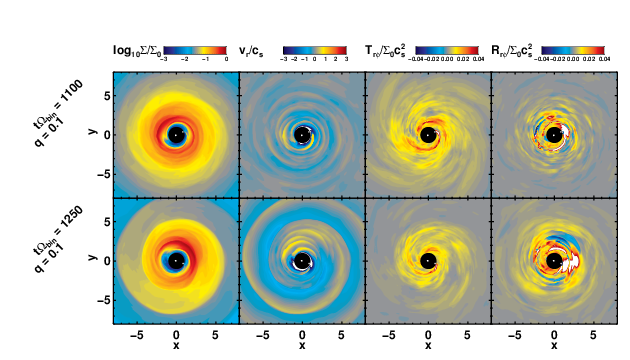}}}
\caption{\small{The same diagnostics as in Figure~\ref{fig:twophase_q=1}, but for the $q=0.1$ disk
		at $t=1100$ and $1250\omgb^{-1}$. In this case, the spiral wave transition took 
		place at $t\sim 1200\omgb^{-1}$, slightly earlier than in the equal mass case. }}
\label{fig:twophase_q=0.1}
\end{figure*}

These two phases of accretion are also closely connected to the phase transition of the disk
structure for the $q=0.1$ run. In Figure~\ref{fig:twophase_q=0.1}, we show snapshots of disk
properties before and after the transition. It is clear that the disk evolves from a relatively
compact two-armed
spiral structure into a large scale single-armed structure. We notice that this transition happens
faster than in the $q=1$ run, possibly due to the asymmetry of the binary system
itself. A common feature shared by these two phases is the gas stream attached to the
secondary, which indicates stronger accretion onto the secondary than onto the primary. 


Like the $q=1$ case, the accretion rate through the inner boundary in the $q=0.1$ case (black solid curve in upper
panel of Figure~\ref{fig:mdot_t_r}) appears to have two phases as
well.  At early times ($t=1050$--$1250\omgb^{-1}$), the time-averaged rate $\dot{M}(r=r_{\rm{in}})\simeq
0.013 (GMa)^{1/2}\Sigma_0$, while the accretion rate gradually increases so that the average at late times
($t=1250$--$1500$) is $\simeq 0.017 (GMa)^{1/2}\Sigma_0$.  In both phases, the time-averaged
$\dot{M}(r=r_{\rm{in}})$ exceeds the $q=1$ case by $\sim 20\%$, and is $\sim 1.5$--$2$ times the
single mass case \footnote{Here we have ignored the effects from the different $r_{\rm in}$ as they are
negligibly small. The accretion rate of the equal mass binary hardly changes between $r=0.8 a$ and
$r=1.0 a$. }.   
By comparison, in the 2D hydrodynamic simulations of \citet{dorazio13}, the time-averaged accretion rate
onto a $q=0.1$ binary was $\simeq 0.7\times$ that of a single-mass system, while those of \citet{farris2014}
found a ratio $\simeq 1.7$.   Just as for the $q=1$ case, our results show a qualitatively similar but quantitatively
somewhat greater accretion efficiency than found by \citet{dorazio13}, and quite good quantitative
agreement with \citet{farris2014}.

\subsection{Accretion rate fluctuations}

In both the $q=1$ and $q=0.1$ disks, the fractional amplitude of fluctuations in the mass accretion rate
is a few times greater than for a point-mass ($q=0$).  As we will argue below, the larger fluctuations are
driven by the binary itself.  However, the accretion rate fluctuations in the binary cases also appear
to depend significantly on mass-ratio, both in amplitude and in frequency. 
The typical peak-to-trough amplitude contrast for $q=0.1$ is $\simeq 0.01$--$0.015$ $\Sigma_0(GMa)^{1/2}$,
about as large as the mean accretion rate.   This is about twice the amplitude seen when $q=1$.

\begin{figure}[t!]
\epsscale{0.7}
\plotone{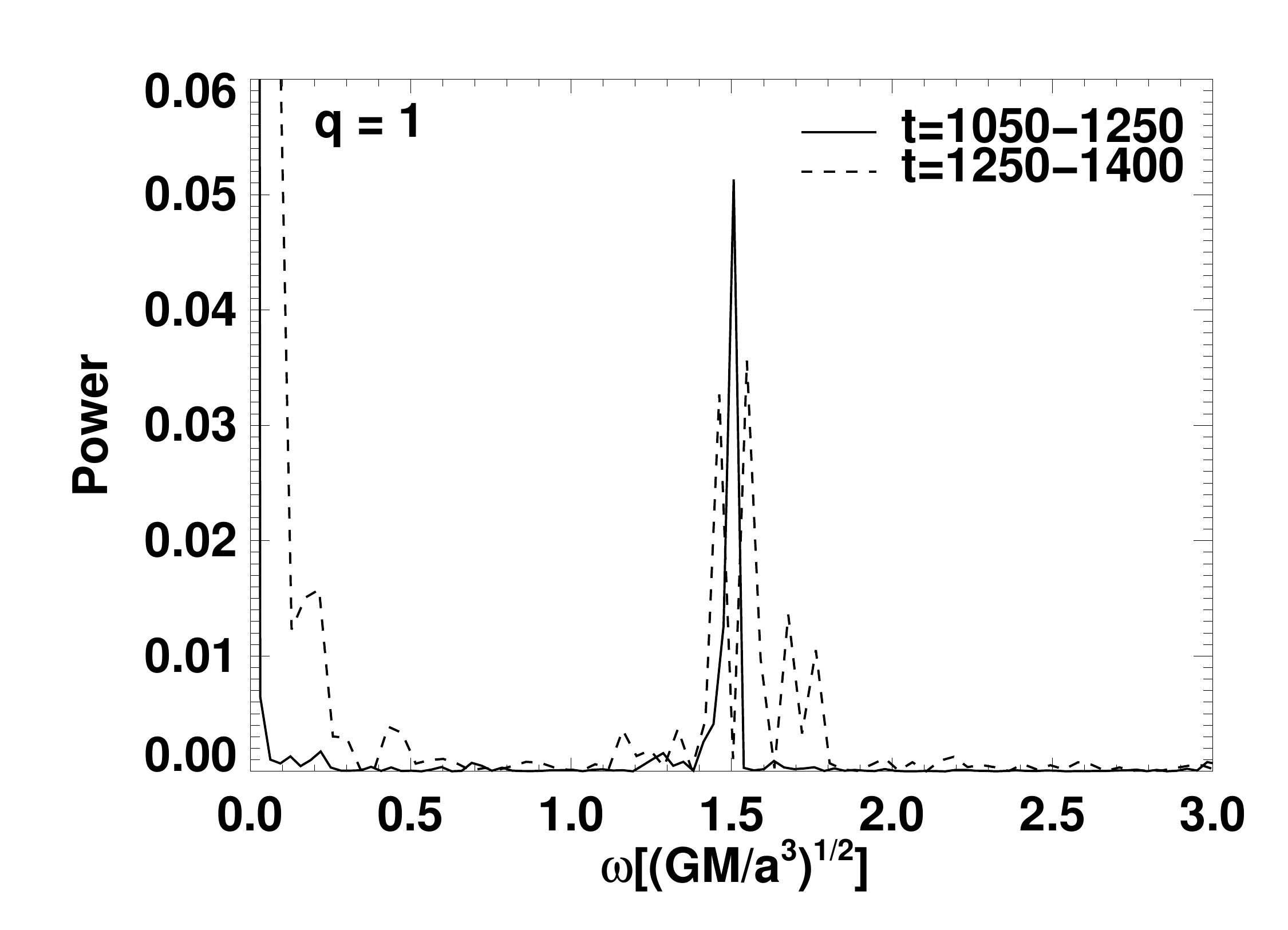}
\plotone{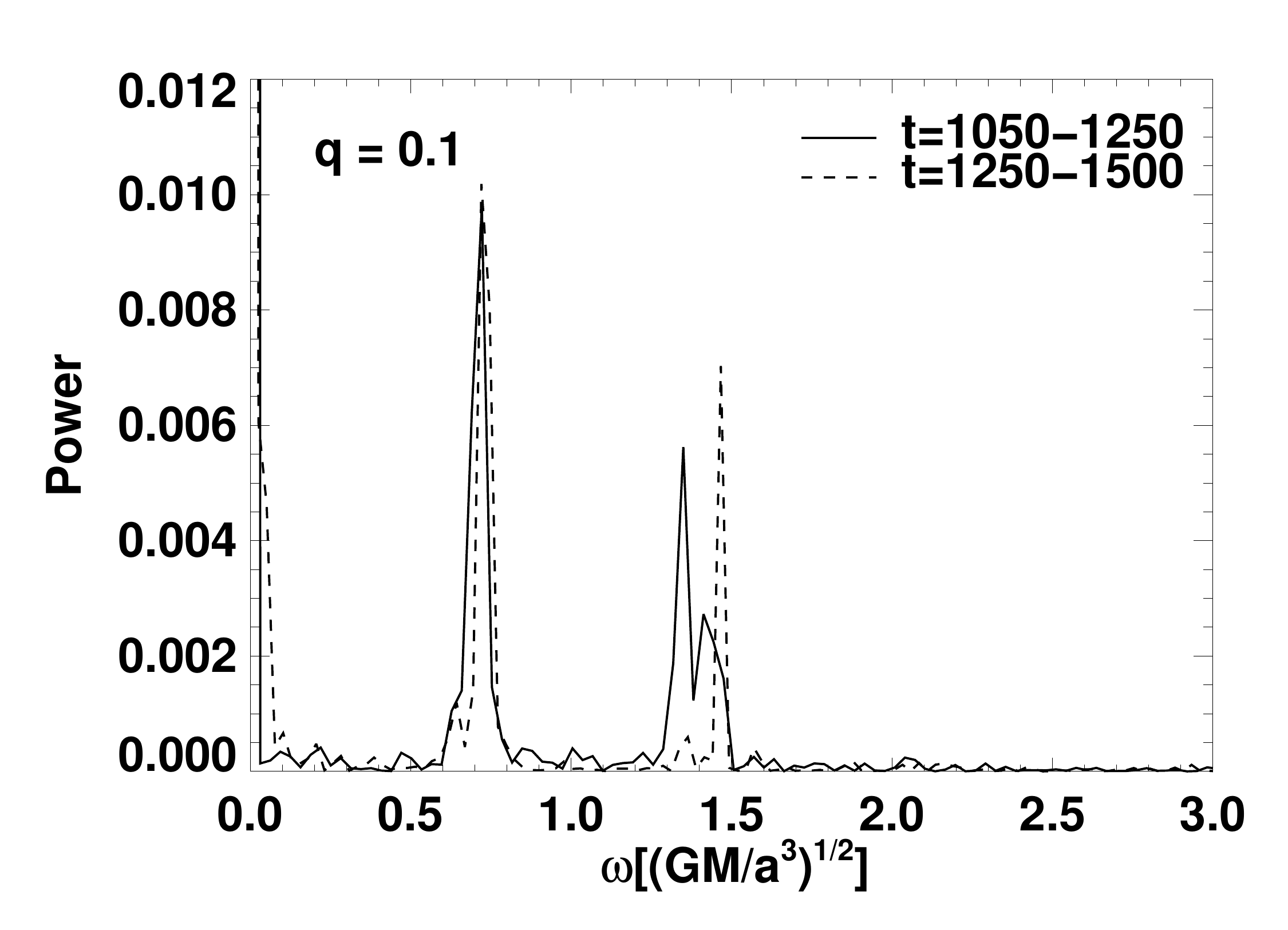}
\caption{\small{Fourier decomposed power spectrum of the accretion through the inner boundary of
		a $q=1$ binary (top panel) and a $q=0.1$ binary (bottom panel).  }}
\label{fig:mdot_pspec}
\end{figure}

At early times in the $q=1$ run, the accretion rate fluctuates periodically at a frequency of $\sim 1.5\omgb$, as
shown in Figure~\ref{fig:mdot_pspec}.   
This is the stage, called the `transient state' by \citet{dorazio13}, in which the disk is beginning to become
elliptical and two streams of nearly equal strength run inward from the disk's inner edge; \citet{dorazio13}
similarly found a periodic modulation with this frequency.    Later in the simulation, the $1.5\omgb$ peak in the
power spectrum splits into two lower magnitude spikes, their frequencies centered on the original peak
frequency but separated by $\simeq 0.1\omgb$. This split indicates a change in the disk structure
from point-symmetric toward more eccentric shape.

Unlike the $q=1$ case, the $q=0.1$ case can induce $m=1$ disk asymmetry immediately.   Consequently,
right from the start we see strong peaks in the accretion rate power spectrum at $\simeq 0.7\omgb$ and
$\simeq 1.3\omgb$.  These modes are beats between the binary frequency $\omgb$ and the orbital
frequency of the small density enhancement near the disk's inner edge, $\simeq 0.3\omgb$.   Another
peak at $\simeq 1.4 \omgb$ might be the first harmonic of the $0.7\omgb$ beat.  At later time, we still
see the $0.7\omgb$ and $1.4\omgb$ beat frequencies, but they shift slightly toward higher frequency as the
disk gap expands by a small amount.   Interestingly, we do not see the strong single peak at $1.0 \omgb$
observed by \cite{dorazio13} for this mass ratio.

\section{Analysis\label{sec:analysis}}
%

Having seen that $\epsilon$ is actually slightly {\it greater} than unity, we now turn to an effort to understand
this result.   The first point to raise is that it is unlikely $\epsilon > 1$ can persist for a long period of time.
If it were to do so, the inner region of the circumbinary disk ($r \gtrsim 2a$) would be drained of mass, inevitably
leading to a reduction in the accretion rate onto the binary.   Thus, the better way to think about the values of
$\epsilon$ seen in our simulations, a few tens of percent greater than unity, is that the spiral waves excited in
a circumbinary disk by the members of the binary create a sufficient enhancement of the Reynolds
stress to raise the accretion rate per unit mass in the inner disk by a few tens of percent.   By this means,
an accretion rate equal to that injected at large radius can be sustained by a surface density somewhat
smaller than required when the potential is due to a point-mass. Over longer times than we can follow
with this kind of simulation, we expect that the surface density in the inner disk will decline to this level,
leaving the disk in true inflow equilibrium.

With that clarification, it is time to consider the question of why 2D and 3D simulations consistently see
substantial accretion from circumbinary disks onto the central binary despite the contrary prediction made
by 1D studies.   One clue to the answer comes from the structure of the accretion flow through the cavity:
narrow streams.

\subsection{Stream Structure\label{sec:explain_accretion}}

\begin{figure}[!t]
\epsscale{0.8}
\plotone{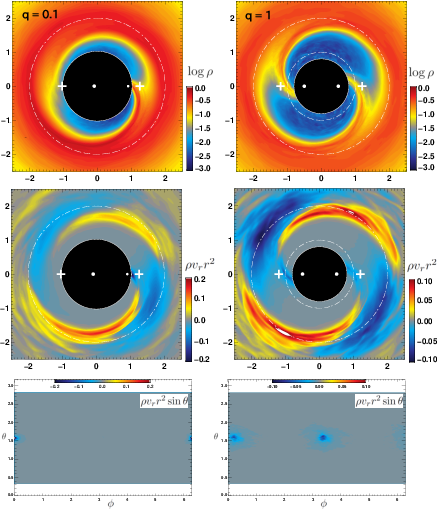}
\caption{\small{Left column: Time averaged midplane density (top), midplane   
		(middle) and inner boundary (bottom) accretion rate $\rho v_r r^2 \sin\theta$,
                both for the $q=0.1$ binary over the last $50\omgb^{-1}$ of the simulation.
                All figures in a frame comoving with the binary.
                Right column: same as left, but for $q=1$ binary. 
		Here negative means inflow. The plus symbols in the midplane plots mark the L2 and L3 points. 
                Summed separately, regions of inward and outward mass flux have comparable
                magnitude; their net, although smaller in magnitude, is consistently inward.
}}
\label{fig:avgmdot}
\end{figure}
In the body of an accretion disk, the inflow speed is generically much slower than the orbital
speed, $\sim \alpha (H/r)^2 v_{\rm orb}$, where $\alpha$ is the usual ratio of vertically-integrated
stress to vertically-integrated pressure, and $H$ is the local scale height.   On the other hand, this
flow, although only $\sim H$ thick in the vertical direction, takes place, on average, around the
entire circumference of the disk, through an area $2\pi r$ wide.

By contrast, the flow across the cavity (see Fig~\ref{fig:avgmdot}) is restricted to very narrow
streams.  Along the central density maximum of the streams, they are typically $\sim 2$--$3H$ wide if
measured sideways from the maximum to where the density drops by $90\%$.
Moreover, the density in the streams as they approach the inner boundary is $\sim 3$--$10$ times
lower than the density in the disk body.
Thus, in order to carry the same mass inflow, the inward velocity in a stream must be
$\sim (r/H)(\rho_{\rm disk}/\rho_{\rm stream})$
larger than the typical disk inflow speed, or $\sim(\rho_{\rm disk}/\rho_{\rm stream}) \alpha (H/r) v_{\rm orb}$.
This condition can be easily achieved because the absence of stable closed orbits within $r \sim 2a$
when $q$ is not too small leads to a characteristic inward speed $\sim 0.3 v_{\rm orb}$,
whereas $(\rho_{\rm disk}/\rho_{\rm stream}) \alpha (H/r) \sim 0.03$--$0.1$ in the
conditions of our simulation, and often smaller in real disks.

Thus, one way of looking at the contrast between the 1D and the 2D/3D results is to observe that
the 1D picture, in which there is no inward motion within $r \sim 2a$, is correct at almost all, but not
quite all, positions.   We explore this idea further in the next subsection.

\subsection{Orbital properties of inward trajectories\label{sec:test_particle}}
\begin{figure*}[!t]
\epsscale{2.1}
\plottwo{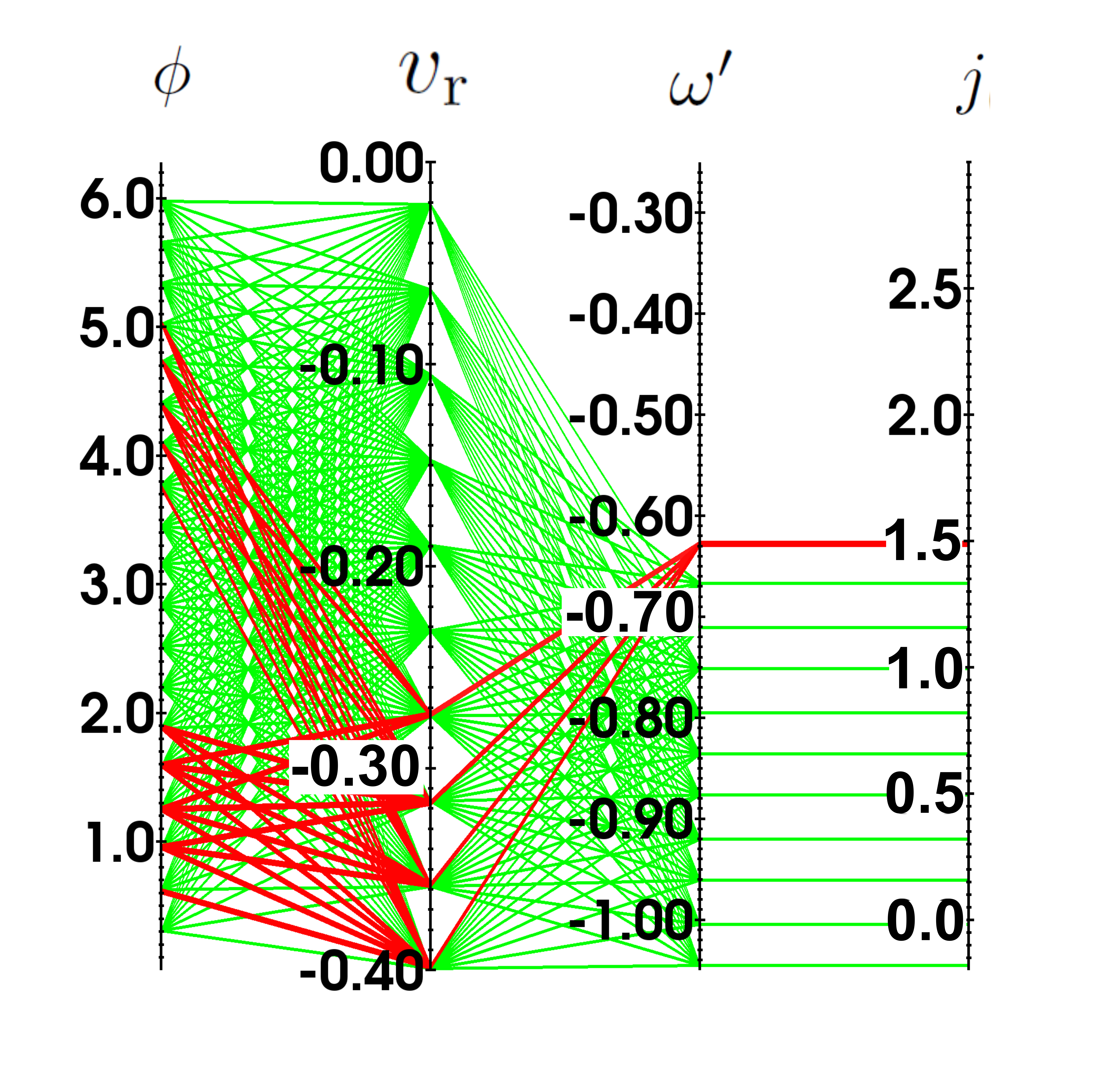}{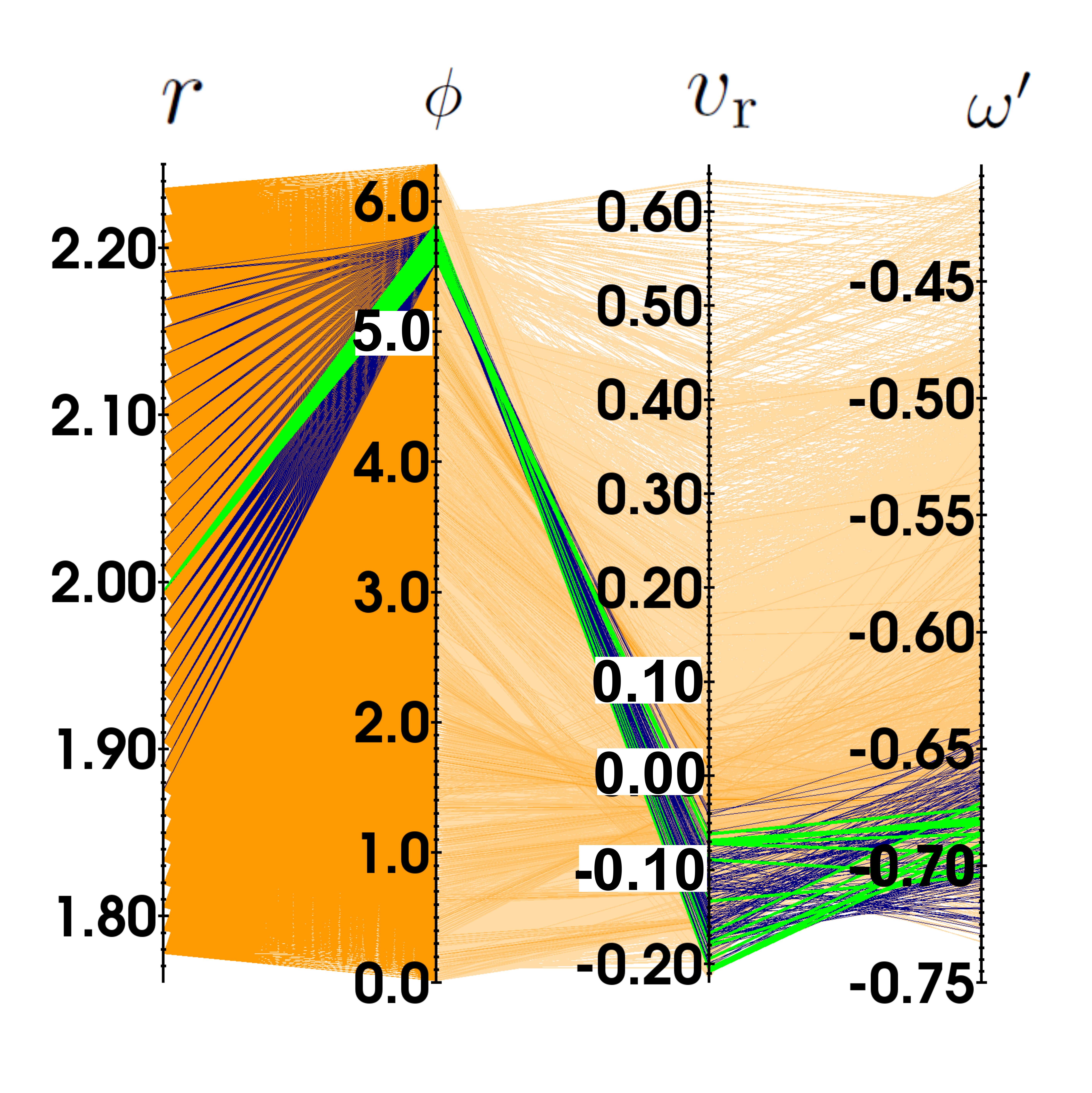}
\caption[caption]{\small{Left: A parallel coordinate plot for test particles at $r=2a$
	distributed uniformly in $\phi$, $v_r$ (shown in units of $(GM/a)^{1/2}$, and
        $\omega^{\prime}=\Omega-\omgb$.  The last coordinate $j\equiv(\omega^{\prime}+\omgb)r^2$ is the
        initial specific angular momentum in units of $(GMa)^{1/2}$.
	The green lines show the initial locations of all infalling particles; note that all have
        $j \lesssim 1.5(GMa)^{1/2}$, the specific angular momentum that supports a circular orbit in the region sampled.
        The red lines show particles with enough angular momentum to trace nearly-circular orbits, but also fall
        inward; all have especially large inward speed as well.
	The red particles concentrate at two opposite azimuth angles where the binary potential is
	relatively weaker than at other angles, making it easier for particles to fall in.
	\\\hspace{\textwidth}
	Right: Similar to the left but now the lines represent individual cells drawn from 
	the actual simulation of the $q=1$ disk at $t=1208$. The orange lines show all cells within
	annulus having $r\simeq 2a$; the blue lines show all the infalling cells; the green lines show
        infalling cells with initial radius exactly $2a$.
	Comparing the orange and blue regions, only a small fraction of the disk cells actually 
	fall in. The green cells found here are consistent in their properties with the green test particles shown on the
	left: $v_r\lesssim 0$ for $\omega^{\prime}\lesssim -0.68$. 
}}
\label{fig:particle}
\end{figure*}

The reason gas accretion occurs through such narrow channels is that only a small fraction
of the gas possesses the proper initial conditions (in position-velocity phase space) for
`infall' trajectories, i.e., trajectories that begin from the disk's inner edge and reach the inner cutoff $r_{\rm in}$. 
Most of the gas near the disk's inner edge that begins moving inward is turned around and
flung back out to the disk by binary torques.   This fate of the majority of the initially inflowing gas
is what led the 1D analysis to predict no net accretion.

To test this explanation and identify exactly what the special conditions for successful inward flow are, we
begin by solving a large number of restricted three-body problems further restricted to orbits entirely in the midplane.
The binary for these trajectory integrations has a circular orbit and $q=1$. 
The test-particles have initial conditions evenly distributed within a phase-space
volume defined by $1.5a \leq  r \leq 2.5a$, $0 \leq \phi \leq 2\pi$, $-0.4\sqrt{GM/a} \leq v_r \leq 0$,
and $-0.4 \leq \omega^{\prime}-[\Omega_{K}(r)-\omgb] \leq +0.4$.
Here $\omega^{\prime}$ is the angular frequency in a frame co-rotating with the binary. 
The initial specific angular momentum is therefore $j\equiv (\omega^{\prime}+\omgb)r^2$.

Our results are shown in the left panel of Figure~\ref{fig:particle} in the form of a Parallel
Coordinates plot\footnote{Parallel Coordinates is a common method for multidimensional
visualization. For our case, each particle is represented by a line connecting multiple attributes
shown as vertical coordinates in parallel.}.   When the initial radius is $r=2a$ (at which a circular
orbit requires $j \simeq 1.45\sqrt{GMa}$ when $q=1$), particles whose initial $j$ is 
$\gtrsim 1.5\sqrt{GMa}$ ($\omega^{\prime}\omgb^{-1}\gtrsim -0.63$) cannot reach the inner boundary; particles
with $1.3 \lesssim j/\sqrt{GMa} \lesssim 1.5$ ($-0.68\lesssim\omega^{\prime}\omgb^{-1}\lesssim -0.63$) can, 
but only if they also have $v_r \lesssim -(0.1$--$0.3) \sqrt{GM/a}$; particles with $j \lesssim 1.3\sqrt{GMa}$ 
are able to travel to $r_{\rm in}$ even with $v_r$ only slightly negative.   The behavior of both these
classes of particles can be understood by reference to the approximate (i.e., ignoring the quadrupolar
contribution) effective potential $V_{\rm eff}(r) \simeq -GM/r + j^2/(2r^2)$ at $r=r_{\rm in}$.    If $v_r^2 \ll V_{\rm eff}(r_{\rm in})$,
the particle radial kinetic energy is negligible, and infall can happen only when $V_{\rm eff}(r_{\rm in}) \leq 0$,
i.e.,  $j \simeq 1.3\sqrt{GMa}$ or less, or equivalently $\omega^{\prime}$ is no more than $\simeq -0.68\omgb$.
Alternatively, if $v_r$ is sufficiently negative, the particle radial kinetic energy can be large enough to overcome a positive
effective potential barrier.

These conditions for infall are not easily met in a real disk because the mean $j$ is close to the circular
orbit value, $\simeq 1.45\sqrt{GMa}$, and radial infall speeds $\lesssim -(0.1$--$0.3) \sqrt{GM/a}$
are rare in the disk body.    Consequently,
only those few fluid elements with $j$ well below the mean can contribute to the inflow, and they are
found in a tightly-constrained portion of phase space. In the right panel of Figure~\ref{fig:particle}, we show a
selection of fluid elements taken from a snapshot of our simulation at $t=1208\omgb^{-1}$.  The samples
are drawn from a ring of disk around $r=2a$ with a width of $0.5a$ in order to capture the inner
edge of the disk. Comparing the orange regions (all cells) with the blue ones (those falling in), one 
can conclude that indeed only a tiny fraction of the disk proper falls to the binary, while the rest of its fluid elements
are pushed back out even if they initially move in a small distance.   The criterion governing which
fluid elements are able to move inward is identical to that identified for the test-particles: $j \lesssim 1.3\sqrt{GMa}$.
In contrast, most of the orange fluid elements have either too large a $j$ (or $\omega^{\prime}$), i.e. large
$V_{\rm eff}$,  or too small an inflow velocity $-v_r$, and are therefore flung out.  Particularly low $j$ fluid elements
move inward so quickly there is too little time for binary torques to substantially raise their angular momentum;
because they can pass the effective potential barrier at $r_{\rm in}$, they cross the inner boundary and
are permanently removed from the circumbinary disk.

Because gas flows near the binary are close to ballistic {\citep{LA2000,shi12}}, we can estimate the
accretion rate by counting the number of fluid elements satisfying the inflow criterion.   The data shown in
the right panel of Figure~\ref{fig:particle} indicate that the blue fluid elements cover $\sim 1/10$ of the annulus
between $r\sim 1.9 a$ and $2.2 a$. For a typical surface density $\sim 0.2\Sigma_0$ and an infall timescale
$\sim 2\pi/\omgb$, the inferred inflow rate would be $\dot{M}\sim 0.012\sqrt{GMa}\Sigma_0$, consistent with our
simulation results.   Thus, although 90\% of inner disk fluid elements cannot go any significant distance inward,
the angular momentum distribution is broad enough that its low $j$ tail suffices to carry the full accretion rate.

Having found that the fluid elements able to accrete are defined by their low specific angular momentum, the
next question to answer is how their angular momentum is reduced to that level.    Ordinary MHD turbulence
does not broaden the angular momentum distribution to this degree: as shown in Fig.~\ref{fig:str_q=1}, the
Reynolds stress in the $q=0$ case is only $\simeq 1\%$ of the pressure, and the pressure is only $\sim 1\%$
of the orbital energy per unit mass.   Instead, the answer appears to be a consequence of the binary
torques themselves.   As previously remarked, most of the mass in the streams that move inward from the
circumbinary disk returns to the disk after its specific angular momentum is raised by those torques.    With
that additional angular momentum, its azimuthal velocity is somewhat greater than the local orbital velocity when
it strikes the disk ($j \simeq 1.6 (GMa)^{1/2}$ as opposed to $j \simeq 1.45 (GMa)^{1/2}$).    The work
done by the torques also increases the streams' energy, giving them an outward radial speed
$\simeq (0.3$--$0.5) (GM/a)^{1/2}$.

\begin{figure}[!h]
\epsscale{1.0}
\plotone{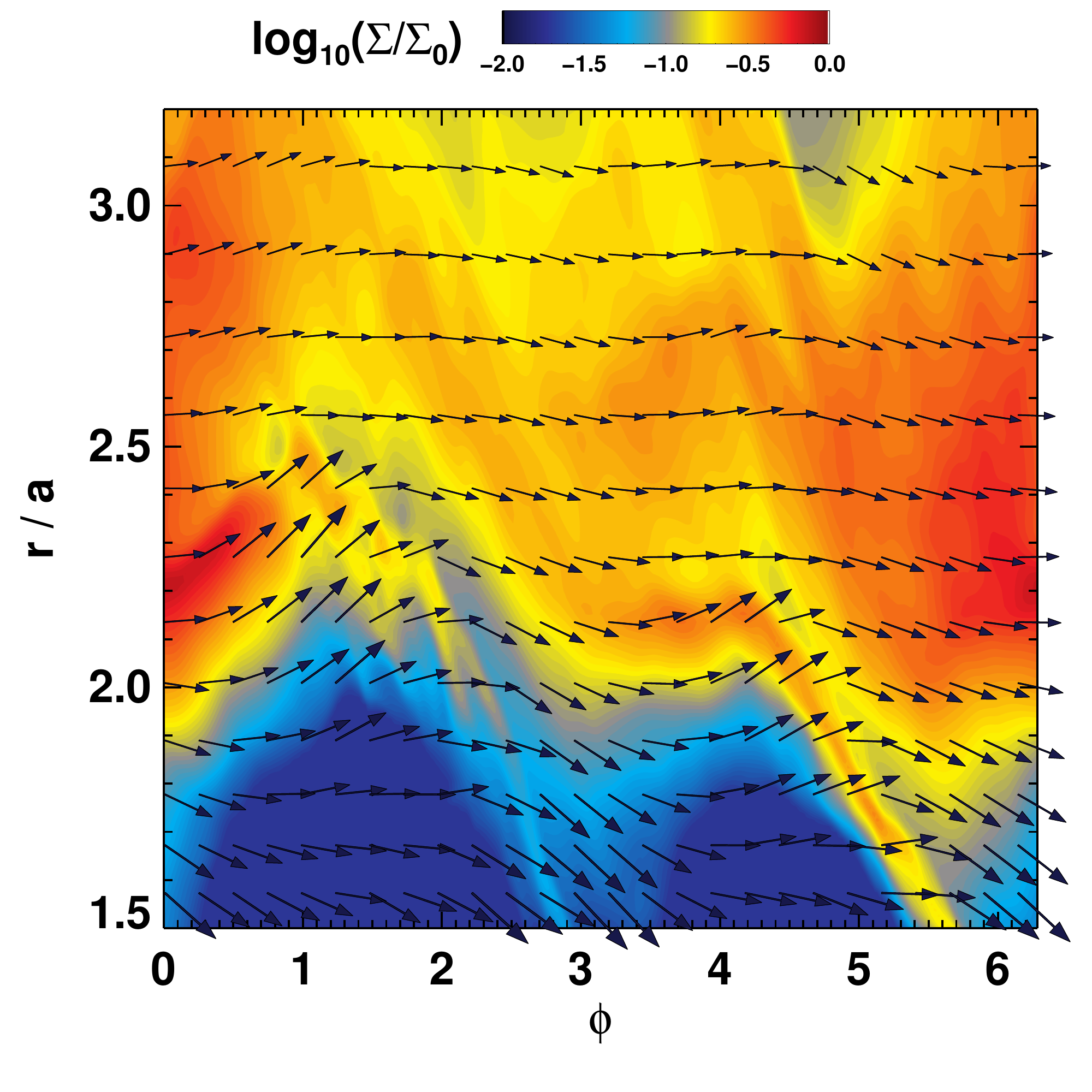}
\caption{\small{Surface density (color contours) and mass-weighted vertically-averaged velocity in the orbital
plane (arrows) at $t=1209.6$.}}
\label{fig:streamshocks}
\end{figure}
An example of such a rapidly-moving outward stream can be seen near $r \sim 2.3$, $\phi \sim 1.5$ in
Figure~\ref{fig:streamshocks}.    Part of the stream ($r \simeq 2.3$--2.4, $\phi \simeq 1$--1.5) is still
heading outward.   However, a part of that stream has already encountered the ridge of high density
trending from larger radius to smaller between $\phi \simeq 2$ and $\phi \simeq 3$.   When it struck
that ridge, a pair of shocks was formed, a forward shock propagating into the disk material
and a reverse shock propagating into the material that had been moving outward.  In the
reverse shock, a portion of the stream mass loses a significant fraction of both its angular momentum
and its orbital energy and heads back inward, seen in this snapshot along the line from ($r \simeq 2.3$, $\phi \simeq 2$)
to ($r \simeq 1.8$, $\phi \simeq 3$).   This shock deflection is the origin of
the low angular momentum material which can now travel all the way to the inner boundary.

Because this mechanism depends almost entirely on 2D orbital mechanics, its nature should be only weakly
dependent on parameters such as the disk's aspect ratio $H/r = c_s/v_{\rm orb}$, provided only that $c_s/v_{\rm orb}$
is small enough that the stream-disk interaction is supersonic.     In our simulation, in which the sound speed is
$0.1(GM/a)^{1/2}$, the Mach numbers of these shocks are $\sim 3$--5; in real disks, they could be considerably
larger.    Support for the view that the stream-disk interaction is at most weakly affected by the disk thickness
also comes from the fact that 2D  simulations employing laminar hydrodynamics and an ``$\alpha$ viscosity" to mock up the
fluid's internal shear stress \citep{farris2014} find a very similar continuity in the accretion rate.   It is possible,
however, that the dynamics of the stream-disk shocks may depend upon the equation of state of the shocked gas.
In our simulation, we assumed that the gas is isothermal.   If, instead, this sort of shock were to occur in a
less lossy gas, the immediate post-shock temperature would be much greater than the disk gas temperature.
The shocked gas could then swell vertically well beyond the thickness of the disk, permitting it to flow
above and below the disk.   Investigation of such effects is well beyond the scope of our effort
here.

\subsection{One-armed Spiral Wave\label{sec:onearm}}
As we have already remarked, a surprising outcome of our simulations is that at late times a strong
one-armed spiral wave propagates outward through the circumbinary disk, enhancing the Reynolds stress
sufficiently to increase the accretion rate by a few tens of percent.   Its origin is worth further attention.

\begin{figure*}[!hb]
\includemovie[controls=true,text={\includegraphics[width=15cm]{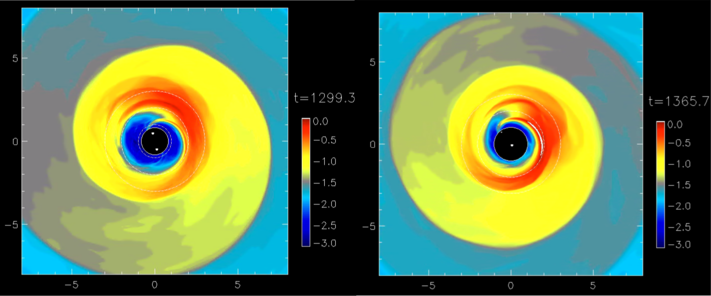}}]
{\linewidth}{0.4\linewidth}{f13.mp4}
\caption{\small{
Streams returning to the circumbinary disk exciting spiral waves. Click the figure to play a short animation
of surface density in the $q=1$ and $0.1$ cases \protect\footnotemark.   
Spiral waves can clearly be seen to begin when a stream pushed outward
by the binary torques strikes the inner edge of the disk.  
}}
\label{fig:spiralwave}
\end{figure*}
\footnotetext{Also available at \url{http://www.astro.princeton.edu/~jmshi/leakage.htm}.}
The most likely cause of this $m=1$ feature is the complement of the mechanisms studied in the
previous subsection.   There we focused on the properties of those special fluid elements able to travel
all the way from the inner edge of the disk to the binary.   Here we focus on all the others,
the ones that may initially move inward, but then feel the torques exerted by the binary and are
propelled outward again until they strike the disk's inner edge. 
To demonstrate how these streams excite spiral waves, we fit the wave form seen in
our simulations (surface density plots in Figure~\ref{fig:twophase_q=1} and \ref{fig:twophase_q=0.1})
with a standard density wave pattern \citep[e.g., Equation~(36) of][]{rafikov2002}. The resulting pattern 
speed is $\simeq 0.17\omgb$ ($\simeq 0.21\omgb$) for the $q=1$ ($q=0.1$) case, which suggests wave 
excitation occurs at $r \simeq 3.2 a$ when $q=1$ and $r \simeq 2.8 a$ when $q=0.1$. 
As shown in Figure~\ref{fig:twophase_q=1} and \ref{fig:twophase_q=0.1}, the density is strongly enhanced
near $r\simeq 3a$.
This connection is most clearly seen in an animation (Fig.~\ref{fig:spiralwave}).

The animations further show that the point of impact, the azimuthal location of the spiral wave driving point, rotates
around the disk's inner edge at an angular frequency slightly lower than the binary frequency.

\begin{figure}[!h]
\plottwo{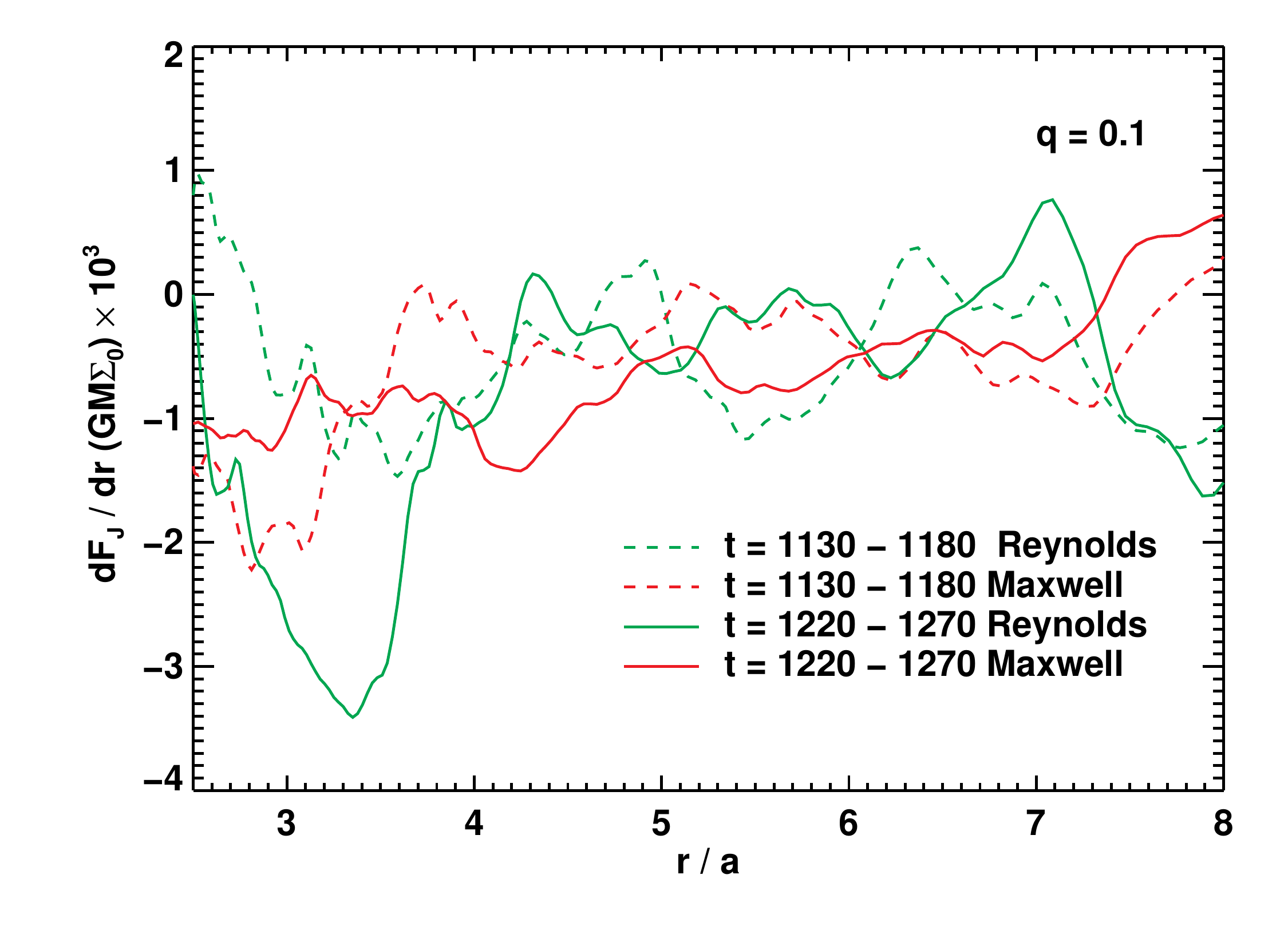}{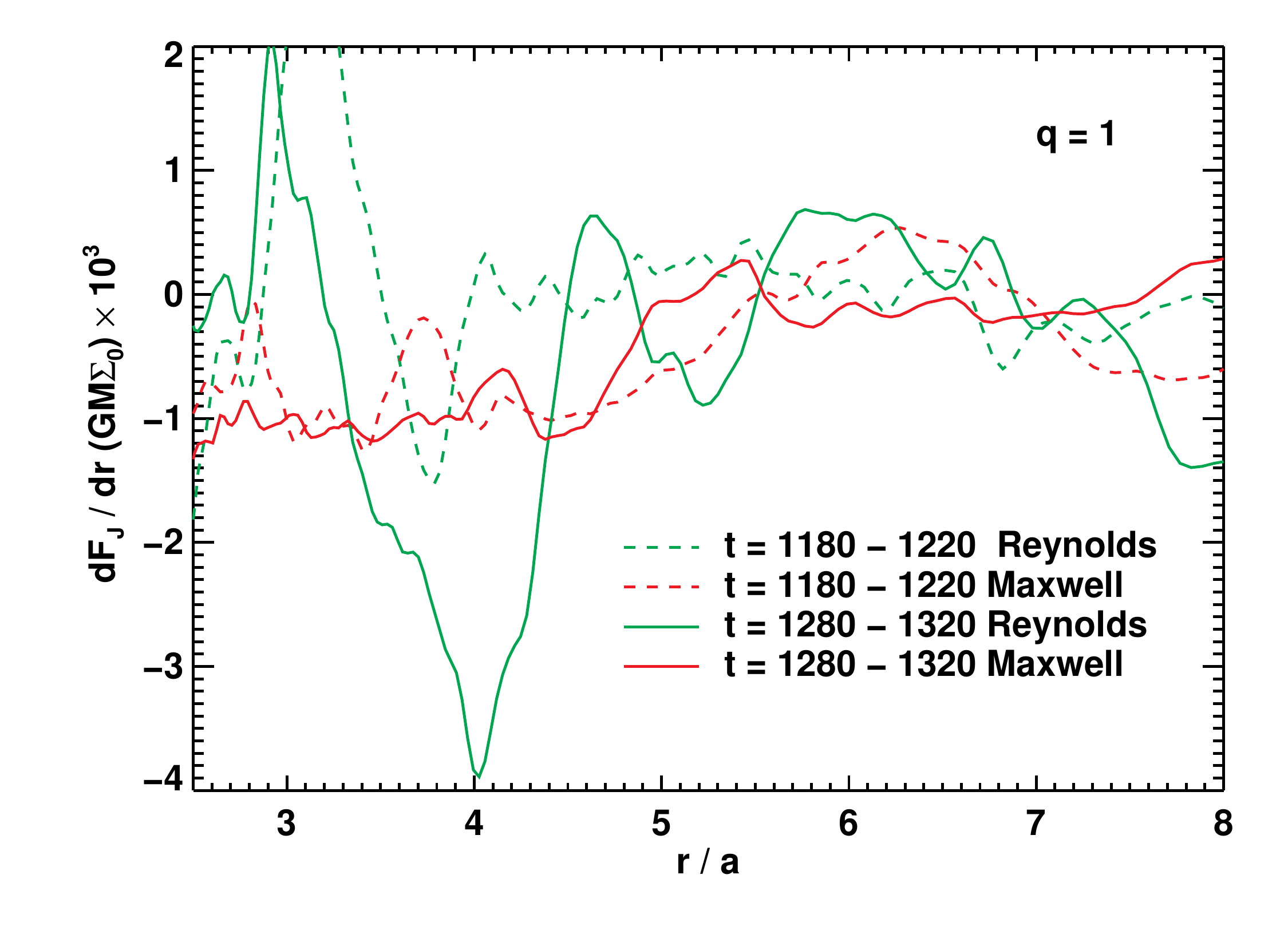}
\caption{\small{ Radial derivatives of time-averaged and shell-integrated angular momentum
flux (AMF) as functions of radius for the $q=0.1$ (left) and $q=1$ (right) cases before and after
the phase transition, where red shows the differentiated Maxwell AMF and green represents
the differentiated Reynolds AMF. Negative means outward transport of angular momentum.
Time averages are long enough that the net change of local angular momentum are close to zero.
The late time averages of the $q=1$ ($q=0.1$) cases show that the Reynolds stress dominates
the angular momentum budget at $r\sim 3$--$4.5a$ ($2.5$--$4a$), a location coinciding with
the stream impact regions shown in Figure~\ref{fig:twophase_q=1} for the $q=1$ case and
Figure~\ref{fig:twophase_q=0.1} for the $q=0.1$ case. }}
\label{fig:amf}
\end{figure}

Similar to tidally or mass-transfer induced spiral shocks that transmit angular momentum outward
in circumstellar disks in a close binary system \citep[e.g.,][]{sawada1986,rs93}, the waves excited 
by the binary-driven streams can also provide long-range angular momentum transport.
Figure~\ref{fig:amf}  shows the net outward angular momentum flux (AMF) associated with the
time averaged Reynolds stress before and after large scale $m=1$ density waves are excited
(see green curves in the bottom two panels). Compared to the early period (top row), the 
Reynolds stress contributes a sizable negative torque at disk radius $\sim 3$--$4.5a$ ($2.5$--$4a$)
at late time (bottom row) in the $q=1$ ($q=0.1$) disk,  which could explain the slightly greater accretion
rate observed in $q\ne0 $ disks than in the $q=0$ disk. 

These large scale one-armed spiral density features were not previously reported in \citet{shi12},
\citet{dorazio13}, or \citet{farris2014}.  We believe they are due to the greater global
isothermal sound speed ($0.1\omgb a$) used here, twice the sound speed in simulation B3D \cite{shi12}.
The faster sound speed causes spiral waves to stretch farther radially, creating
large-scale loosely-wrapped density features \citep{SPL1994}.
In \citet{dorazio13} (as well as in \citep{mm08}), $H/r=0.1$ at all radii, so that the isothermal sound
speed drops $\propto r^{-1/2}$. The wavelength becomes shorter as the wave moves outward,
causing the wave to become more and more tightly wrapped as it travels to greater radius.   Given
this apparent sensitivity to the disk equation of state, it remains to be seen whether such strong spiral
waves are generic or rare in real circumbinary disks.

\section{CONCLUSIONS\label{sec:conclusion}}
In this paper, we carried out 3D global MHD simulations of circumbinary disks with binary mass
ratios $0.1$ and $1$ and contrasted them with a disk orbiting a solitary point-mass.   We found
two major results:
\begin{enumerate}
\item {The time-averaged accretion rate from a circumbinary disk with either $q=1$ or $0.1$ is
indistinguishable from that of a circum-solitary disk whose central mass is the same.  In other words,
essentially all the mass supply given the disk at large radius ultimately leaves its inner edge and travels
to the binary.    The similarity of the $q=1$ and $q=0.1$ cases in this regard suggests that this result
depends at most weakly on binary mass-ratio.   This result confirms, with physical internal fluid stresses,
the conclusion reached by \cite{farris2014} on the basis of 2-d hydrodynamics and a phenomenological
viscous stress.}
\item{The key reason why initial 1D analyses suggested that the accretion efficiency (the parameter
we call $\epsilon$) is $\simeq 0$ rather than $\simeq 1$ is that they omitted consideration of the
small volume in orbital phase space from which trajectories can travel inward from the disk's inner edge
all the way to the binary, avoiding the strong torques that, for most of phase space, push matter back
outward.   Only those fluid elements with specific angular momentum $\simeq 15\%$ less than the
circular orbit value at the disk's inner edge can cross the gap and reach the binary.   Ironically, fluid
elements with exactly this property are created as a consequence of the binary torques themselves:
these torques add enough angular momentum to other gas traveling through the gap to propel it
back out to the disk; it shocks upon reaching the disk, and a portion of its mass is deflected onto
accretion orbits.}
\end{enumerate}

Our results have several observational implications.   If all the matter supplied to a circumbinary
disk at large radius accretes onto the binary through
narrow streams, those streams must shock when they strike the outer edges of accretion disks
around the members of the binary.   In the context of the formation of stellar binaries, these shocks
may be the sites of the $v = 1$--$0~S(1)$ H$_2$ vibrational lines that can often be detected
\citep{beck2012}. 
Radial inflow streams at velocities approaching free-fall can also potentially explain both
the observed low density cavities in transitional disk systems and the relatively normal stellar
accretion rates they maintain \citep{Rosenfeld2014}.   In the context of supermassive black
hole binaries, one can similarly expect strong shocks where the streams strike the outer edges
of the individual disks.   Hard X-rays from those shocks may be observable when the binary
separation is small enough \citep{roedig14,farris2015}.    If the accretion flow fed in at large
radius is large enough, still greater luminosity can be generated when the accreted matter
approaches the two black holes' ISCO regions.

\acknowledgments
We thank an anonymous referee for constructive questions that led to improvement of this paper. 
We also thank Eugene Chiang and Neal Turner for useful discussions.
This research was supported by an allocation of advanced computing resources provided by the
National Science Foundation. The computations were performed on Kraken at the
National Institute for Computational Sciences (http://www.nics.tennessee.edu/).
This work was partially supported by NSF grant AST-1028111 (JHK) and
NASA Origins grant NNX13AI57G (J-M S).

\clearpage

\end{document}